\documentclass[sigconf]{acmart}

\AtBeginDocument{%
  \providecommand\BibTeX{{%
    \normalfont B\kern-0.5em{\scshape i\kern-0.25em b}\kern-0.8em\TeX}}}

\copyrightyear{2022}
\acmYear{2022}
\setcopyright{acmlicensed}
\acmConference[SIGMOD '22]{Proceedings of the 2022 International Conference on Management of Data}{June 12--17, 2022}{Philadelphia, PA, USA}
\acmBooktitle{Proceedings of the 2022 International Conference on Management of Data (SIGMOD '22), June 12--17, 2022, Philadelphia, PA, USA}
\acmPrice{15.00}
\acmDOI{10.1145/3514221.3517825}
\acmISBN{978-1-4503-9249-5/22/06}

\usepackage{hyperref}
\usepackage{listings}
\usepackage{cleveref}
\usepackage{subcaption}
\usepackage{xspace}
\usepackage{enumitem}
\usepackage{multicol}

\begin{document}
\fancyhead{}

\newcommand{\system}{P4DB}
\newcommand{\pfour}{P4}
\newcommand{\naive}{na\"{\i}ve\xspace}
\newcommand{\Naive}{Na\"{\i}ve\xspace}
\newcommand{\naively}{na\"{\i}vely\xspace}

\definecolor{comments}{HTML}{228b22}
\definecolor{background}{HTML}{EEEEEE}
\lstdefinelanguage{p4}{
  sensitive = true,
  backgroundcolor=\color{background},
  keywords={Register, RegisterAction},
  keywords=[2]{if, else, default},
  otherkeywords={\#define, \#undef, \#if, \#else, \#endif, \#ifdef, \#ifndef, \#elif, \#include},
  keywords=[3]{action, apply, bit, const, control, default, enum, error, extern, false, header,
      header_union, in, inout, int, package, parser, out, select, state, struct, table,
      transition, true, typedef, varbit, verify, metadata, header_type, fields, Switch, Pipeline
    },
  keywordstyle=\color{purple}\bfseries,
  keywordstyle=[2]\color{red}\bfseries,
  keywordstyle=[3]\color{blue}\bfseries,
  identifierstyle=\color{black},
  sensitive=false,
  comment=[l]{//},
  morecomment=[s]{/*}{*/},
  commentstyle=\color{comments}\ttfamily,
  stringstyle=\color{red}\ttfamily,
  morestring=[b]',
  morestring=[b]",
}

\lstset{
  language=p4,
  extendedchars=true,
  basicstyle=\scriptsize\ttfamily, %
  showstringspaces=false,
  showspaces=false,
  numbers=left,
  numberstyle=\scriptsize, %
  numbersep=12pt,
  frame=single,
  tabsize=2,
  breaklines=true,
  showtabs=false,
  captionpos=b
}

\title{\system{} - The Case for In-Network OLTP \protect\\ (Extended Technical Report)}

\setcopyright{none}

\author{Matthias Jasny}
\affiliation{%
\institution{Technical University of Darmstadt}
\country{}}

\author{Lasse Thostrup}
\affiliation{%
\institution{Technical University of Darmstadt}
\country{}}

\author{Tobias Ziegler}
\affiliation{%
\institution{Technical University of Darmstadt}
\country{}}

\author{Carsten Binnig}
\affiliation{%
\institution{Technical University of Darmstadt}
\country{}}

\renewcommand{\shortauthors}{Jasny et al.}

\begin{abstract}
  
In this paper we present a new approach for distributed DBMSs called \system{}, that uses a programmable switch to accelerate OLTP workloads.
The main idea of \system{} is that it implements a transaction processing engine on top of a P4-programmable switch.
The switch can thus act as an accelerator in the network, especially when it is used to store and process hot (contended) tuples on the switch.
In our experiments, we show that \system{} hence provides significant benefits compared to traditional DBMS architectures and can achieve a speedup of up to $8\times$.

\end{abstract}

\settopmatter{printacmref=false}
\renewcommand\footnotetextcopyrightpermission[1]{}

\maketitle

\section{Introduction}
\label{sec:introduction}

\paragraph{Motivation} %
The efficient use of data center networks plays a significant role on the performance of distributed DBMSs.
Traditionally, distributed DBMSs were built on the assumption that the network is a major bottleneck. As such, classical distributed DBMSs were designed to mitigate the effects of the high network cost using sophisticated techniques such as complicated partitioning schemes \cite{quamar2013sword,pavlo2012skew,curino2010schism,zamanian2015locality}, semi-join transformations \cite{michael2007improving,polychroniou2014track,rodiger2016flow}, speculative execution \cite{pavlo2011predictive}, new consistency levels \cite{kraska2009consistency}, or even the relaxation of atomicity guarantees \cite{krishnaswamy1997relative,levy1991theory}.
However, data center networks have been evolving significantly in recent years.

A first major trend that we have seen in the last years is that data center networks have evolved \emph{from being slow to being fast}.
For example, when looking at the network speed provided by cloud vendors such as Amazon, Microsoft or Google for their hosted cloud instances, we see that even for the smaller (i.e., cost-efficient) instances, the network link has a speed of at least 10Gbps and can reach up to 100Gbps for larger instance types.
Consequently, there are significant efforts in database community into investigating how these high-speed networks can be used efficiently by redesigning distributed DBMSs for zero-copy protocols such as RDMA or DPDK to avoid the overhead of classical network stacks such as TCP/IP \cite{zamanian2019rethinking,ziegler2019designing,binnig2016end,zamanian2016end,kalia2016fasst,wei2015fast}.

Another major trend that we have seen in the recent years is that with the rise of software-defined networks \cite{kreutz2014software}, switches and network cards have become programmable \cite{ports2019should,bosshart2013forwarding} and thus turned networks \emph{from being passive to being active}.
This programmability of the network opens up many additional opportunities to tailor the network to the applications on top.
In particular, the programmability of the so-called data plane allows applications to offload computation to the network devices (aka in-network-processing or INP for short).
INP has shown to provide significant benefits for distributed data processing in general, including key-value stores \cite{yu2020netlock,li2016fast,jin2018netchain} as well as distributed OLAP and ML \cite{tirmazi2020cheetah,sapio2021scaling,blocher2018boosting,lerner2019case,li2019accelerating}.
However, as far as OLTP is concerned, there has been only limited work so far which offloads only certain sub-components of OLTP, such as lock management \cite{yu2020netlock} or replication protocols \cite{jin2018netchain,zhu2019harmonia}.

\begin{figure}
    \captionsetup[subfigure]{aboveskip=0.0ex,belowskip=0.0ex}
    \captionsetup{aboveskip=0.0ex,belowskip=-0.0ex}
    \begin{subfigure}{.64\columnwidth}
        \centering
        \includegraphics[width=0.99\linewidth]{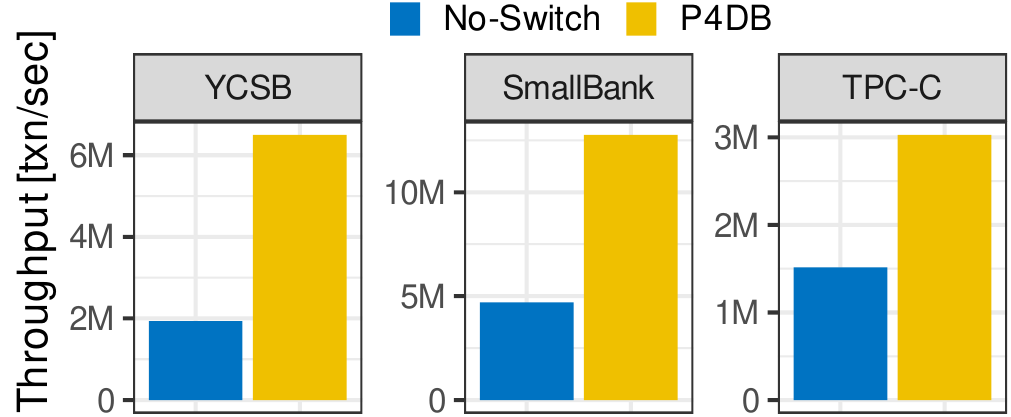}
        \caption{Throughput}
        \label{fig:frontpage_throughput}
    \end{subfigure}
    \begin{subfigure}{.35\columnwidth}
        \centering
        \includegraphics[width=0.99\linewidth]{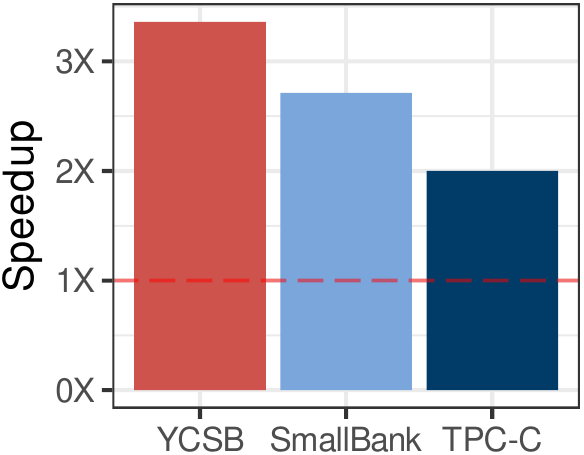}
        \caption{Speedup}
        \label{fig:frontpage_speedup}
    \end{subfigure}
    \caption{OLTP processing in \system{} using a programmable switch. Compared to a traditional DBMS without using the switch, \system{} provides significant speedups. Details about the setup are explained in our evaluation.}
    \label{fig:frontpage}
    \vspace{-4.5ex}
\end{figure}

\vspace{-2.0ex}\paragraph{Contributions} As a main contribution, in this paper we present a new, more aggressive approach for distributed DBMSs called \system{} that involves a programmable switch more actively to accelerate OLTP workloads.
The main idea of \system{} is that it implements a \emph{full transaction processing engine on top of a programmable switch}, using P4 as the de facto data plane programming language.
That way, \system{} exposes the switch as an ``additional'' database node that, however, comes with very different properties compared to a normal database node (as we discuss below).
As a result, the switch is an ideal ``place'' to store and process hot (contended) tuples that typically lead to a significant performance degradation in traditional DBMSs.
As we show in \Cref{fig:frontpage}, \system{} can thus significantly speed up OLTP processing in distributed DBMSs for various workloads in case hot tuples are offloaded to the switch.

To better understand where the benefits of using a switch originate from, let us first look at the hardware characteristics of (programmable) switches that make them an interesting candidate for hot tuples.
As a main difference to a normal database node of the host DBMS, a programmable switch comes with two interesting characteristics: (1) Programmable switches are designed to process the aggregated load of network traffic from all connected servers at line-rate. (2) The switch can be reached from database nodes that are directly connected to the switch with only half of the network latency compared to the latency required to reach any other (remote) database node that is connected to the same switch.
Based on these characteristics, we next argue that a programmable switch is an ideal place for storing and processing hot items of an OLTP workload that are most frequently accessed by transactions.

In a traditional distributed DBMS architecture, hot items that are accessed frequently by transactions typically lead to a severely degraded performance, as mentioned before.
This is especially true, when hot items are involved in distributed transactions, as hot items typically suffer from increased remote access latencies.
The latency increase, in turn, leads to an increased likelihood of contention on those items and thus higher abort rates \cite{zamanian2020chiller}.
In contrast, \system{} can mitigate these effects if hot items are stored and processed on a programmable switch: %
This is because when hot tuples are involved in distributed transactions, storing them on a switch reduces access latencies, which reduces the overall contention, directly leading to performance benefits.

Surprisingly, however, as we show in our evaluation, even if an OLTP workload is perfectly partitionable and no distributed transactions are involved in a workload, or even if there is only limited skew in a workload, \system{} can still speed up processing compared to a traditional DBMS.
The reason is that programmable switches provide an execution model that enables a new lock-free execution scheme for transactions, which allows the switch in \system{} to process transactions on hot tuples at high speeds since it avoids any contention.
However, implementing a transaction engine on top of a programmable switch does not come ``for free'' and many challenges need to be addressed.
For example, typical programmable switches of today, such as recent Tofino-based switches \cite{bf2556x}, have restrictions not only on the memory model (i.e., how data in the switch memory can be accessed by in-switch-programs), but also what kinds of in-switch-programs are supported by their execution models.
By making clever use of the switch memory and the execution models of programmable switches though, we can efficiently support the concurrent execution of transactions in an abort-free manner on the switch, as mentioned before.

Finally, we think \system{} is deployable and compatible with existing datacenter networks.
As a main deployment model, \system{} targets cloud providers that have dedicated racks for database services, which is a common scheme to provide high performance for distributed DBMSs \cite{yu2020netlock}.
For such a deployment, \system{} only needs to augment the Top-of-Rack (ToR) switches with a custom data plane module for processing transactions on hot tuples.
But other deployments, e.g. hierarchies of switches or multi-tenant deployments, are clearly also interesting.
While such deployments are out of the scope of this paper and are important avenues of future work, at the end of the paper we provide a short discussion of how we think \system{} can generalize to such deployments.

\vspace{-1.5ex}\paragraph{Outline} The remainder of this paper is organized as follows:
First, in Section \ref{sec:background} we discuss the relevant background on programmable switches and their memory and execution models.
In Section \ref{sec:system_overview} we provide an overview of \system{} before we then explain in Section \ref{sec:data_storage} and Section \ref{sec:txn_processing} the details of the storage and execution model we implemented to execute database transactions on a programmable switch.
Afterwards, in Section \ref{sec:integration} we then discuss the integration of the in-switch transaction processing into a host DBMS. %
Finally, we conclude with an extensive evaluation of \system{} in Section \ref{sec:evaluation} using various OLTP benchmarks as well as an overview of related work in Section \ref{sec:related_work} and a summary in Section \ref{sec:conclusion}.

\section{Background}
\label{sec:background}

In the following, we discuss the relevant background on the typical programmable switches that are commercially available today.

\vspace{-1.5ex}
\subsection{Programmable Switches}
\label{sec:background_switches}

Programmable switches are an emerging trend on the market with specialized ASICs from vendors such as Intel \cite{inteltofino} or Cavium \cite{cavium}. Compared to traditional switches, programmable switches provide capabilities for flexible packet processing at line-rate of up to billion pkt/s.
In-network processing (INP) enables advancements in the programmability of the data plane by providing a reconfigurable architecture \cite{bosshart2013forwarding} based on match-action tables. This increases the flexibility of switches to be used not only for data routing but also for offloading application logic to the switches by implementing customer match-action rules.

The de facto standard for programming the data plane \cite{bosshart2014p4} is P4 (\underline{P}rogramming \underline{P}rotocol-Independent \underline{P}acket \underline{P}rocessors) .
P4 is a high-level language that allows to write match-action rules to express the processing of packets in the data plane \cite{bosshart2014p4}.
Initially designed for programmable network switches, P4 is now used in a variety of systems which can also be used to process packets including SmartNICs or FPGAs \cite{barefootwhitepaper,kundel2019p4,wang2017p4fpga}.

P4 programs in general operate on packet headers and specify how to rewrite those headers.
Although it uses a C-like syntax, it does not allow the use of pointers and has many other restrictions regarding floating-point numbers or loops to allow the processing of P4 programs at line-rate. These constraints can be overcome by common techniques like loop-unrolling, fixed-point arithmetic or dictionary encoding for strings. Switch vendors can also add different dedicated hardware-accelerators (e.g.: a checksum-engine or FPU) to the ASIC, which are then available through \pfour{}-externs.

\begin{figure}
    \captionsetup{aboveskip=0.0ex,belowskip=0.0ex}
    \centering
    \includegraphics[width=.99\columnwidth]{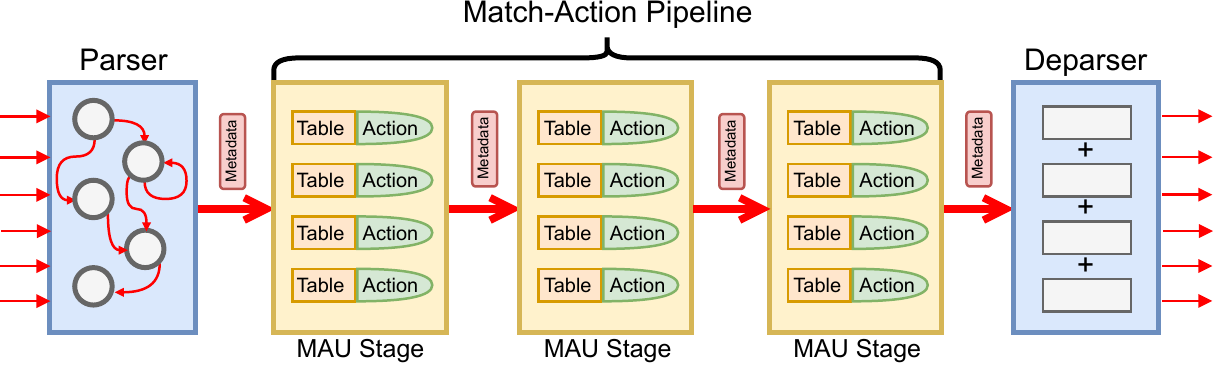}
    \caption{The Protocol Independent Switch Architecture (PISA): All stages (Parser, MAU Stages and Deparser) are programmable and allow flexible packet processing in the data plane based on packet headers and metadata.}
    \label{fig:pisa}
    \vspace{-4.5ex}
\end{figure}

\vspace{-1.0ex}
\subsection{Protocol Independent Switch Architecture}
For executing P4 programs, the commercially available program\-mable switches typically implement the so-called Protocol Independent Switch Architecture (PISA) as shown in \Cref{fig:pisa}.
PISA provides protocol-independence by allowing programmers to specify how a packet should be parsed and processed in a declarative manner using match-action tables as mentioned before.
In \system{}, for example, we use custom match-action rules to implement the logic of the transaction engine on the switch.

To process such application-specific match-action rules, PISA-based switches apply the following procedure: First, the header of a network packet is parsed in the ingress stage. These packet headers can include typical network information that is used for routing the packet along with metadata that can trigger more application-specific match-action rules that are executed in so-called match-action-units (MAUs) of PISA-based switches.
In a PISA-based switch, multiple of these MAUs are typically linked together to process network packets in a pipeline-parallel manner with one packet per MAU stage. An important aspect is that match-action rules, if they depend on each other, must be placed in successive MAU stages. At the end of a switch pipeline, a packet is deparsed and sent over the network to its recipient.

\vspace{-1.5ex}
\subsection{Tofino Native Architecture}
\label{sec:background_tna}

In this paper, we use a Tofino-based switch which is one of the commercially most successful programmable switches today.
Based architecturally on PISA, it provides several extensions as part of the so-called Tofino Native Architecture (TNA) and programmable fixed-function components. For example, switch ports can be configured as loopback ports to cycle the same packet multiple times through the switch. This can be used to implement loops and thus execute data plane programs that can not be executed in a single pipeline pass.
Stateful operations are also an interesting extension available in TNA \cite{kannan2019precise,budiu2017p416}. While these features are Tofino-specific, similar functions are provided by other vendors as well.

The ability of executing stateful operations in switches is a crucial part in \system{}, since it allows us to do transaction processing in a switch.
The idea of these stateful operations in general is that they allow match-action rules to access state in SRAM through so-called register arrays.
However, the total amount of memory, in today's switches is limited to a few megabytes only. %
While this seems to be a quite limiting factor, such an amount of memory is already enough to hold data for a few thousands of hot tuples, which has proven to be sufficient to significantly increase throughput of applications utilizing programmable switches \cite{jin2017netcache,jin2018netchain,yu2020netlock,li2016fast}. For example, the switch-programs of \system{} can store approximately 820K 8Byte hot tuples per pipeline. Moreover, there already exist other switch implementations based on FPGAs and high-bandwidth-memory that further alleviate this limitation \cite{hofmann2019high}. %

Besides the limited switch memory, there exist other constraints on how data can be accessed from within P4 programs that are important aspects we need to consider when designing \system{}.
For example, a first and important constraint, is that the register arrays (i.e., the SRAM) are partitioned over MAU stages and thus can only be accessed when the packet is processed by that stage.
Moreover, to ensure the processing of packets at line-rate, multiple accesses (e.g., multiple read and write operations) from one packet to the very same register are not allowed \cite{yu2020netlock}.
Finally, a last but important limitation of stateful operations is that the order of register accesses in a P4 program needs to follow the placement order over the MAU stages in a pipeline (i.e., a data item stored in a register in an earlier MAU stage must be accessed first in the P4 program).

\section{System Overview}
\label{sec:system_overview}

\begin{figure}
    \captionsetup{aboveskip=0.0ex,belowskip=0.0ex}
    \centering
    \includegraphics[width=.99\columnwidth]{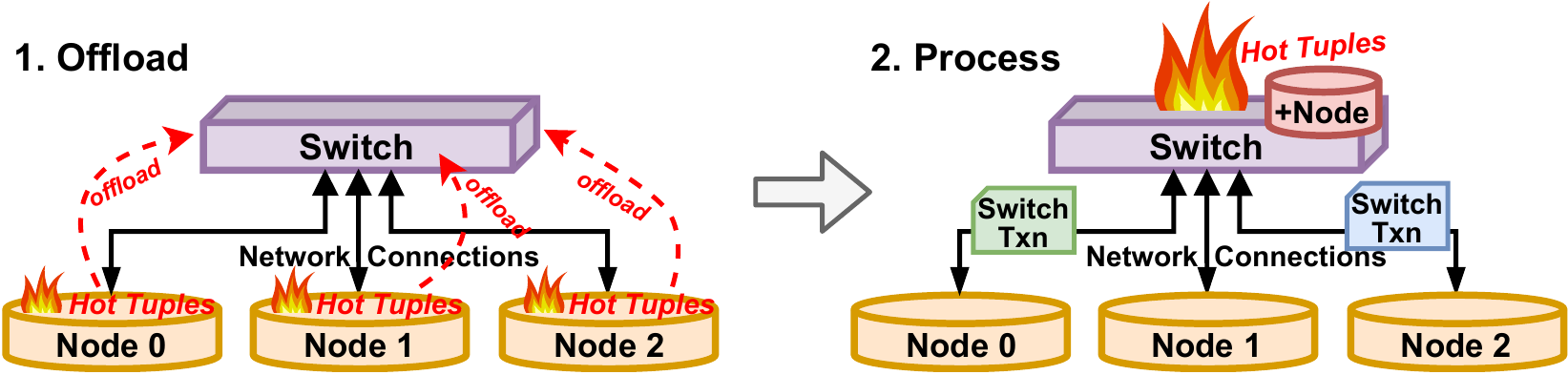}
    \caption{\system{} accelerates distributed OLTP by exposing a programmable switch as an additional database node. To make use of the switch, \system{} first offloads hot tuples to the switch in an offline step. At runtime, transactions can then be executed at line-rate on these hot items on the switch.}
    \vspace{-3.5ex}
    \label{fig:switch}
\end{figure}

In the following, we give an overview of \system{}. As shown in \Cref{fig:switch}, in an offline preparation step (\Cref{fig:switch}, left-hand-side) hot tuples are offloaded from database nodes to the switch. Once offloaded, at runtime (\Cref{fig:switch}, right-hand-side) database nodes of the host DBMS then trigger transactions on the switch.

\subsection{Offloading Data}
A key challenge when offloading hot tuples to a programmable switch is to determine the data layout for the hot tuples on the switch.
Storing tuples in registers of PISA-like switches such as the Tofino and supporting stateful operations on top comes with several constraints as discussed before.
More precisely, hot tuples that should be stored on a switch have to be (1) assigned to registers in the order of how they are being accessed by transactions, and (2) a switch register that stores a tuple can only be accessed once per transaction and pipeline pass.
In case these constraints can be satisfied by the data layout, transactions can be executed on the switch in a single pass through the pipeline.
In Section \ref{sec:data_storage} we hence present a new storage scheme for PISA-like switches called the declustered storage model, that aims to find a data layout for a given set of hot tuples and a set of transactions to maximize the number of transactions that can be executed in a single pass.
Otherwise, if this is not the case, they require multiple passes through the switch pipeline, which negatively impacts the overall performance of executing switch transactions, as we discuss next.

In \system{}, we currently use a static approach to offload data; i.e., we decide based on access statistics of a representative workload which tuples are hot and thus should be offloaded to the switch. 
For the offline detection of hot tuples, we replay the transactions in a workload  statement-by-statement to identify frequently accessed tuples.
To identify those hot tuples at runtime, the partitioning manager of \system{}, which resides on each database node, keeps an index with the tuple identifiers (i.e., their primary keys) as well as other metadata.
More details how this index is used at runtime and how secondary index lookups of hot tuples are supported is discussed in \Cref{sec:integration}.
Clearly, one could also use a more dynamic approach which monitors access frequencies at runtime to support potential shifts in the workload.
However, this is an orthogonal aspect for future work.

\subsection{Processing Transactions}

Once hot tuples are offloaded to a switch and the data layout (i.e., a mapping of tuples to registers) is defined, database transactions can be executed on the switch.
In the following, we assume the case that database transactions can be partitioned into hot/cold transactions that either involve only hot tuples (on the switch) or cold tuples (on the host database nodes). Transactions that span over cold and hot tuples (called warm transactions) will be discussed in Section \ref{sec:integration}.

\vspace{-1.5ex}\paragraph{Hot Transactions}
Hot transactions can be fully executed as in-switch transactions without coordinating with database nodes.
The transaction logic of hot transactions is implemented as P4-programs in \system{} and deployed together with the data in the offloading step.
For executing a hot (in-switch) transaction at runtime, database nodes of the host DBMS need to send a network message to the switch.
This message includes all relevant transaction parameters such as transaction type that should be executed along with its input parameters.
Once a hot transaction is executed, the results are sent back to the calling database node over the network.

The novelty of \system{} lies in the fact how transactions are being processed on the switch.
The details this scheme will be discussed in \Cref{sec:txn_processing}.
In a nutshell, our scheme enables that hot transactions (in most cases) can be executed using a single pass of the transaction through a switch pipeline.
An important aspect is that single-pass transactions can be executed at line-rate and do not require any coordination on the switch while still guaranteeing serializability.
In case transactions can not be executed in a single pass (which can be the case for more complex transactions with many reads and writes), \system{} falls back to a scheme for those transactions that requires multiple circulations through the switch pipeline and relies on coordination (as we discuss later). Hence, a goal of our storage scheme that we discuss next is to lay out the data on the switch to minimize the need for multi-pass transactions.

\vspace{-1.5ex}\paragraph{Cold Transactions} Different from hot transactions, cold transactions are executed completely by database nodes of the host DBMS without involving the switch.
For processing transactions on cold tuples, not stored on the switch, \system{} currently uses a standard 2PC protocol for distributed transactions (with different deadlock detection modes) and a pessimistic lock based concurrency scheme to coordinate execution on each node.
However, it is important to note that any other concurrency scheme could be used to execute cold transactions in the host DBMS (cf. \Cref{sec:integration}).

\section{Declustered Storage Model}
\label{sec:data_storage}

In this section, we present our new declustered storage model for PISA-based switches like the Tofino switch.
The goal of the declustered storage model is to execute hot transactions in a single pass through a switch pipeline.

\subsection{Goals of Data Layout}

In the following, we first provide a high-level overview of how hot transactions are being executed in a switch.
Afterwards, we derive the design of the data layout to enable single-pass transactions on the switch (i.e., which tuples are assigned to registers of the different MAU stages in a switch).

\begin{figure}
    \captionsetup{aboveskip=0.0ex,belowskip=0.0ex}
    \centering
    \includegraphics[width=.99\columnwidth, trim=0mm 0mm 0mm 0mm, clip]{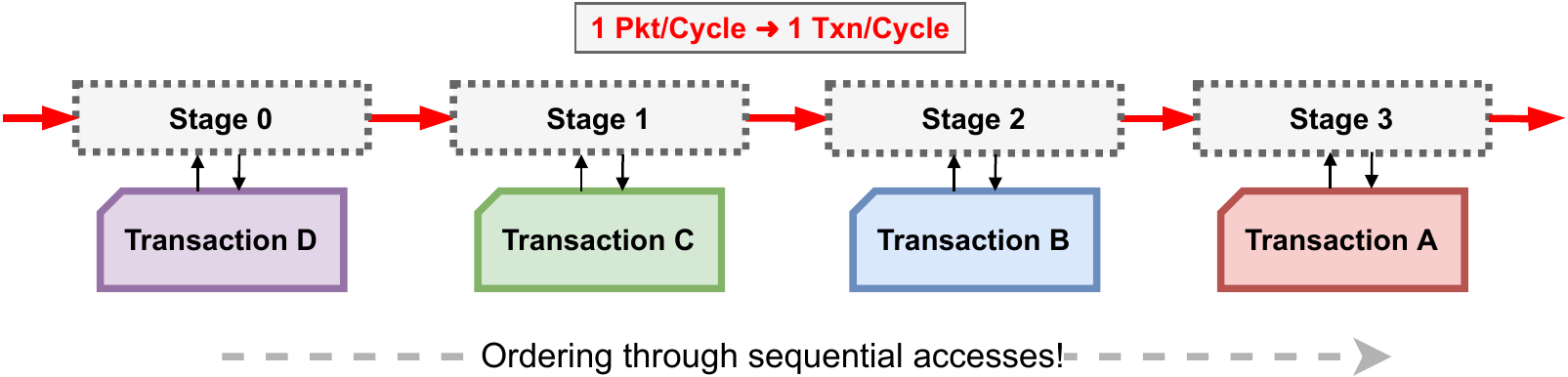}
    \caption{Pipelined processing of switch transactions. Packets represent transactions which advance in the pipeline once per clock cycle. In each cycle, a transaction can only access a single tuple of the current stage with one operation. }
    \label{fig:txn_pipeline}
\end{figure}

As shown in \Cref{fig:txn_pipeline}, in \system{} each network packet in a switch pipeline represents a separate transaction.
This assumption is reasonable, since (1) switches can execute billion packets per second, and (2) batching is often undesired in latency-critical OLTP.
As a result, in each MAU stage of a switch pipeline, there is only a single transaction per cycle.
Moreover, during processing, at each clock cycle transactions move to the next (subsequent) stage in a pipeline which is dictated by the switch execution model of PISA-based switches.

A first important implication that this execution model has on the data layout is that we only have one transaction per MAU stage with exclusive access to the registers of this MAU stage.
This simplifies concurrency control for transactions on the switch, as we discuss next in \Cref{sec:txn_processing}. The first important constraint that is dictated by the switch memory model though is that a transaction can execute only one operation (read or write) to a register index which holds a single tuple. A stage usually can hold a few register arrays.
Consequently, since transactions typically involve read/write operations on multiple tuples, a goal of the switch data layout is to allocate the tuples accessed by the same transaction into distinct register arrays of MAU stages. If this is not possible, a transaction needs to pass through the switch pipeline multiple times to execute all its required (read/write) operations.

Second, as mentioned before in \Cref{sec:background}, the access order of transactions has an influence on the assignment of tuples to registers of MAU stages.
For example, if in \Cref{fig:txn_pipeline} \textit{Transaction B} reads a tuple $A$ and writes a tuple $B$ using $B=B+A$, then $A$ has to be located in a MAU stage before tuple $B$.
Consequently, another goal of the switch data layout is to align the tuples stored in the register arrays of MAU stages with the access order imposed by transactions.
Again, if it is not possible to reflect the order of accesses by a transaction in the data layout, a hot transaction needs to be executed using multiple passes through the switch pipeline.

Finally, there are hot transactions which always require multiple passes independent of the data layout (i.e., a single-pass execution is not possible at all). This is the case, when a transaction needs to execute multiple operations on the same tuple (e.g., a read and later a write of the same tuple), since only one operation per stage to the same tuple is allowed by the memory order of the switch.
However, transactions that require multiple accesses to the same tuple are rather an exceptional case, as we see in our evaluation. %
Furthermore, as we will see in our experiments, \system{} can then still achieve significant benefits if such a case occurs compared to a traditional DBMS architecture without a switch.

\subsection{Data Layout Problem}
\label{sec:data_layout_problem}

The goal of the declustered storage model is (as discussed before) to enable that hot transactions can be executed in a single pass by assigning tuples to switch registers that is optimal for the switch workload.
More precisely, given a set of hot tuples and a set of hot transactions that should be executed on the switch, the goal of the declustered data layout is to maximize the number of hot transactions in the workload that can be executed in a single pass.
Interestingly, this problem is related to declustered data access for disk-based databases that aims to spread data access over different disks  \cite{liu1996partitioning,liu1995similarity,liu2001hypergraph}.

\begin{figure}
    \captionsetup{aboveskip=0.0ex,belowskip=0.0ex}
    \centering
    \includegraphics[width=.99\columnwidth, trim=0mm 0mm 0mm 0mm, clip]{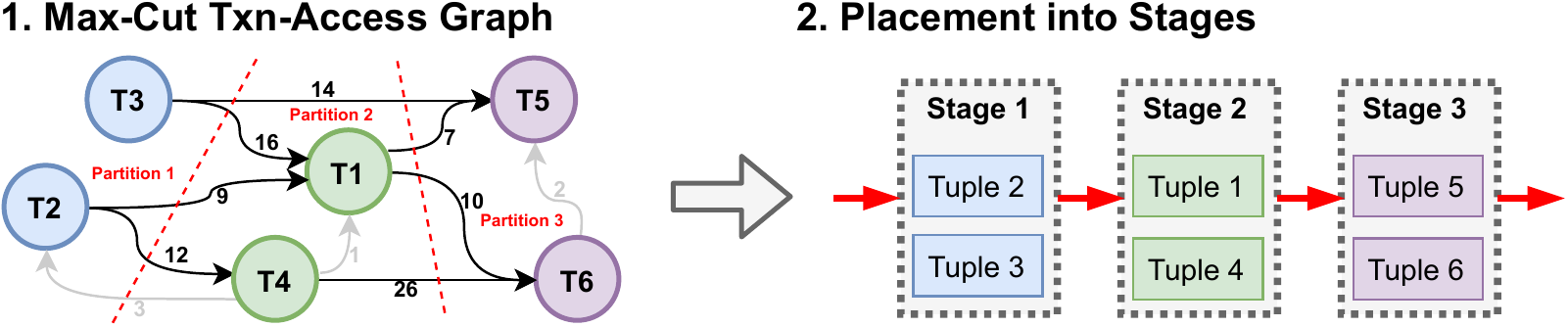}
    \caption{Placement of tuples into switch stages and registers using the data layout algorithm on a graph representation. In a first step (left-hand-side), a max-cut is applied to the graph. Then, in a second step (right-hand-side), the partitions are then assigned to registers of different stages.}
    \label{fig:maxcut}
    \vspace{-3.5ex}
\end{figure}

The problem of declustering data accesses is typically formalized as a graph problem.
The general idea is to represent all tuples for which we aim to define a declustered data layout as nodes in the graph.
If two tuples, represented by nodes \textit{U} and \textit{V}, are accessed together by one transaction, a new edge \textit{e(U,V)} is added to the graph with a corresponding weight that describes their relative access frequency.
In \system{}, we extend this model by using directed edges instead of edges without direction in order to model dependent operations that impose an order of access and thus an order of how to assign tuples to registers.
More precisely, if an ordering dependency of operations (e.g., a read-write dependency) between two tuples represented by nodes \textit{U} and \textit{V} exist in one transaction, the edge is directed from \textit{U} to \textit{V}.
If no such dependency exists, then a bidirectional edge is used between \textit{U} and \textit{V} to indicate that the tuples have no dependencies in the access order. %

\subsection{Data Layout Algorithm}

The data layout algorithm of \system{} uses this graph to assign hot tuples to registers of different MAU stages, with the goal to maximize the number of single-pass transactions.
Recall from before, an edge \textit{e(U,V)} with a high weight in the graph means that two tuples represented by the nodes are frequently accessed within the same transaction. Therefore, they should be placed into register arrays of separate MAU stages.
In the following, we first ignore the edge directions and include them later in the discussion.

Overall, to find an optimal assignment of hot tuples to registers, we can apply a max-cut to partition nodes into $N$ disjoint sub-graphs that we aim to assign to $N$ register arrays of the different MAU stages in a switch pipeline.
The resulting partitions represent the set of tuples that should be allocated to the same register array of a stage.
However, the max-cut problem of a graph is NP-complete \cite{karp1972reducibility}. %
Fortunately, many sophisticated approximation algorithms for the max-cut problem exist \cite{dunning2018works,garey1974some,sahni1976p,kim2019comparison,goemans1995improved,seo2014edge}.
For solving the max-cut problem of finding a partitioning of the hot-set on the switch in \system{}, we use MQLib \cite{dunning2018works} with additional constraints on the maximum size per partition, such that all tuples in a partition fit into the registers of a stage.
\Cref{fig:maxcut} (left-hand-side) shows an example of a graph that was constructed for six tuples (\textit{T1} to \textit{T6}) with weights representing access frequencies and the resulting max-cut for this graph into $3$ different partitions.

After applying the max-cut, in the next step of the data layout algorithm, we need to assign each resulting partition to a register of the switch stage.
For doing this, we now take directions of edges in the cut into consideration.
In case the edges in a cut are either bidirectional edges or if all unidirectional edges in the cut between two partitions point into the same direction, then we assign the partitions with the outgoing edges into an earlier switch stage.
For example, in \Cref{fig:maxcut} (right-hand-side) this is the case for the two cuts that separate the Partition 1 (blue) from Partition 2 (green) and Partition 2 (green) from Partition 3 (purple).
In this case, we can simply do a topological ordering of the partitions based on the direction of the edges in the cuts and assign the partitions in the resulting order to switch stages.

In case that the edges in a cut point into different directions, we apply a slightly modified procedure.
In this case, we first remove those edges in a cut such that all point in one direction and the sum of their weights is smaller than for the edges in the same cut that point in the other direction.
The main idea is that these edges represent the accesses for which we have to violate the access order, resulting in a multi-pass transaction to support this order. Hence, we remove those edges with the total lower access frequency.
The edges that remain in the cut between two partitions now all point to the same direction, and thus we can apply the procedure as discussed before to assign partitions to switch registers.

\begin{figure}
    \captionsetup{aboveskip=0.0ex,belowskip=0.0ex}
    \centering
    \includegraphics[width=.99\columnwidth]{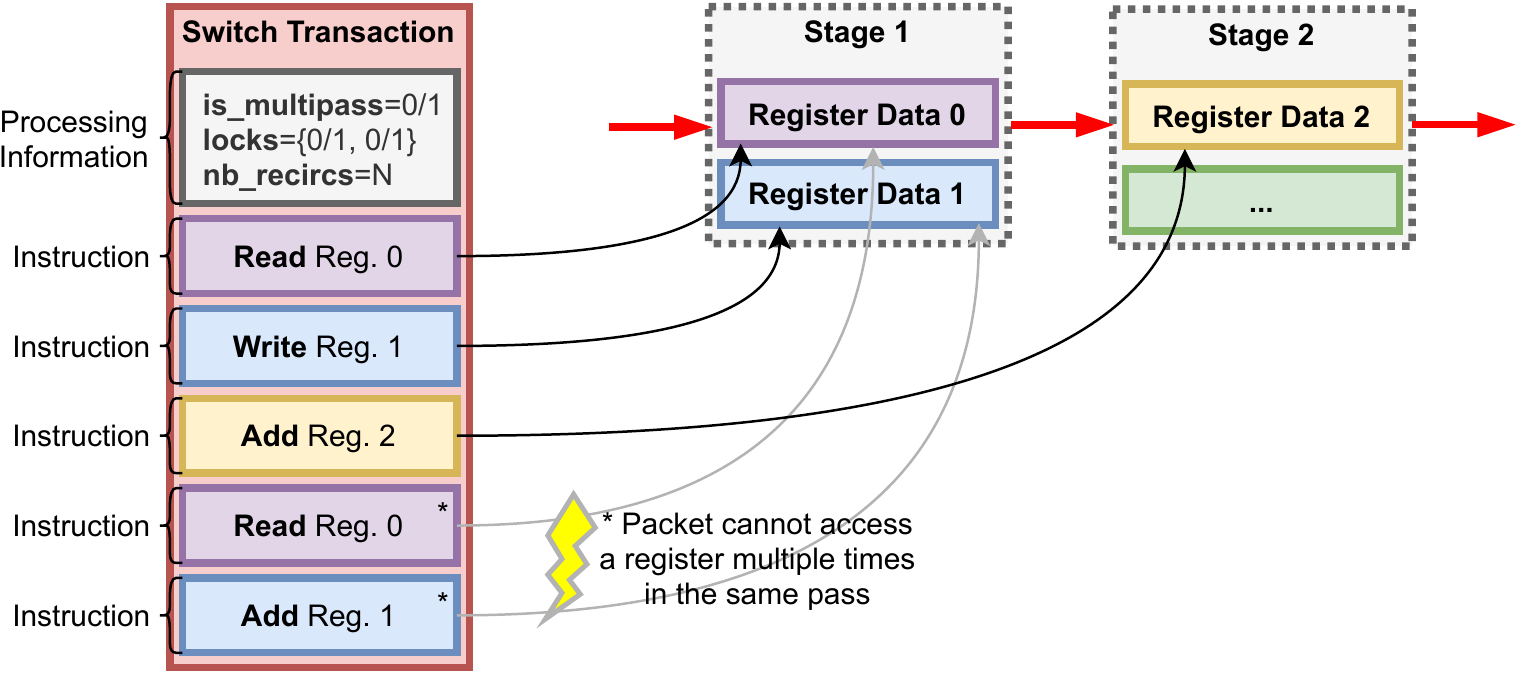}
    \caption{Packet format of switch transactions. Each switch transaction corresponds to one network packet with a header and a variable amount of instructions, each of which defines an operation of a transaction.}
    \label{fig:switch_instr}
    \vspace{-4.5ex}
\end{figure}

\section{In-Switch Transaction Processing}
\label{sec:txn_processing}

In this section, we describe how hot transactions are executed on the switch.
We first explain the base case where a switch transaction can be executed in a single pass, before we then discuss transactions that might require multiple passes.
Overall, the execution of hot transactions on the switch provides guarantees for \textit{Atomicity}, \textit{Isolation} and \textit{Consistency} as we discuss next, while \textit{Durability} is guaranteed by the integration into the host DBMS (cf. \Cref{sec:integration}).

\subsection{Single-Pass Transactions}

In \system{}, as discussed in \Cref{sec:data_storage}, we use a model where one transaction is mapped to one network packet.
An interesting aspect of this model is that we get isolation and atomicity of transactions on the switch without additional coordination in case a transaction can be executed in a single pass.

First, we describe why our switch model actually guarantees isolation out-of-the-box.
To better understand why this is the case, we refer back to \Cref{fig:txn_pipeline}, which shows the pipelined execution of transactions in a switch pipeline.
The important aspect is that our model dictates that there is only one transaction per MAU stage and transactions in different MAU stages are not re-ordered. %
Hence, the pipelined execution of transactions on the switch is equivalent to a serial execution order of transactions that results from the order as they are routed through the pipeline.
For example, in \Cref{fig:txn_pipeline} the execution is equivalent to the serial order: \textit{A}, \textit{B}, \textit{C}, and then \textit{D}.

Second, another interesting aspect that results from the execution model in \Cref{fig:txn_pipeline} is that it also guarantees atomicity out-of-the-box, since no lock conflicts or deadlocks can occur on the switch.
However, constraint checks to enable consistency could potentially still lead to aborts and thus would require additional efforts on the switch to roll back a transaction in case constraints are violated.
To avoid these additional overheads for single-pass transactions, we use so-called constrained-writes of P4 to implement constraint checks.
The main idea of a constrained-write in P4 is that a write is only executed if a predicate is satisfied.
This allows us to support simple constraint checks at the end of a single-pass transaction on individual tuples, such as writing a transaction balance in SmallBank only if its value is larger than zero at the end of the execution.
For supporting more complex constraint checks though, one needs to fall back to a multi-pass execution scheme.

\vspace{-1.5ex}
\subsection{Multi-Pass Transactions}

Hot transactions that are executed on a switch do not always satisfy the restrictions for single-pass execution.
For example, \Cref{fig:switch_instr} shows a transaction that can not be executed in a single pass.
As we can see, this transaction contains multiple operations, while the last two operations access the same register as the first two operations. Hence, this switch transaction needs to be executed in two passes. In the first pass, the three instructions at the beginning are executed and in the second pass the batch of the two remaining instructions.
Yet, if the coordination-free execution would be used for multi-pass transactions, the isolation of transactions is no longer guaranteed.

To support isolation for multi-pass transactions, we thus provide a lock-based execution scheme in \system{}.
In the following, we explain the \naive{} (fallback) approach that can always be used for multi-pass transactions and discuss further optimizations subsequently.
The main idea of processing multi-pass transactions in the \naive{} scheme is that we introduce a so-called pipeline-lock to prevent the concurrent execution of multiple transactions in the same pipeline.
The pipeline-lock is located in the first MAU stage of a pipeline.
When a new transaction arrives at the switch pipeline and another multi-pass transaction is currently in the pipeline, the lock prevents its execution. This transaction is then scheduled for another pass through the switch using recirculation of the network packet.

\begin{figure}
    \captionsetup{aboveskip=0.0ex,belowskip=0.0ex}
    \centering
    \includegraphics[width=.99\columnwidth]{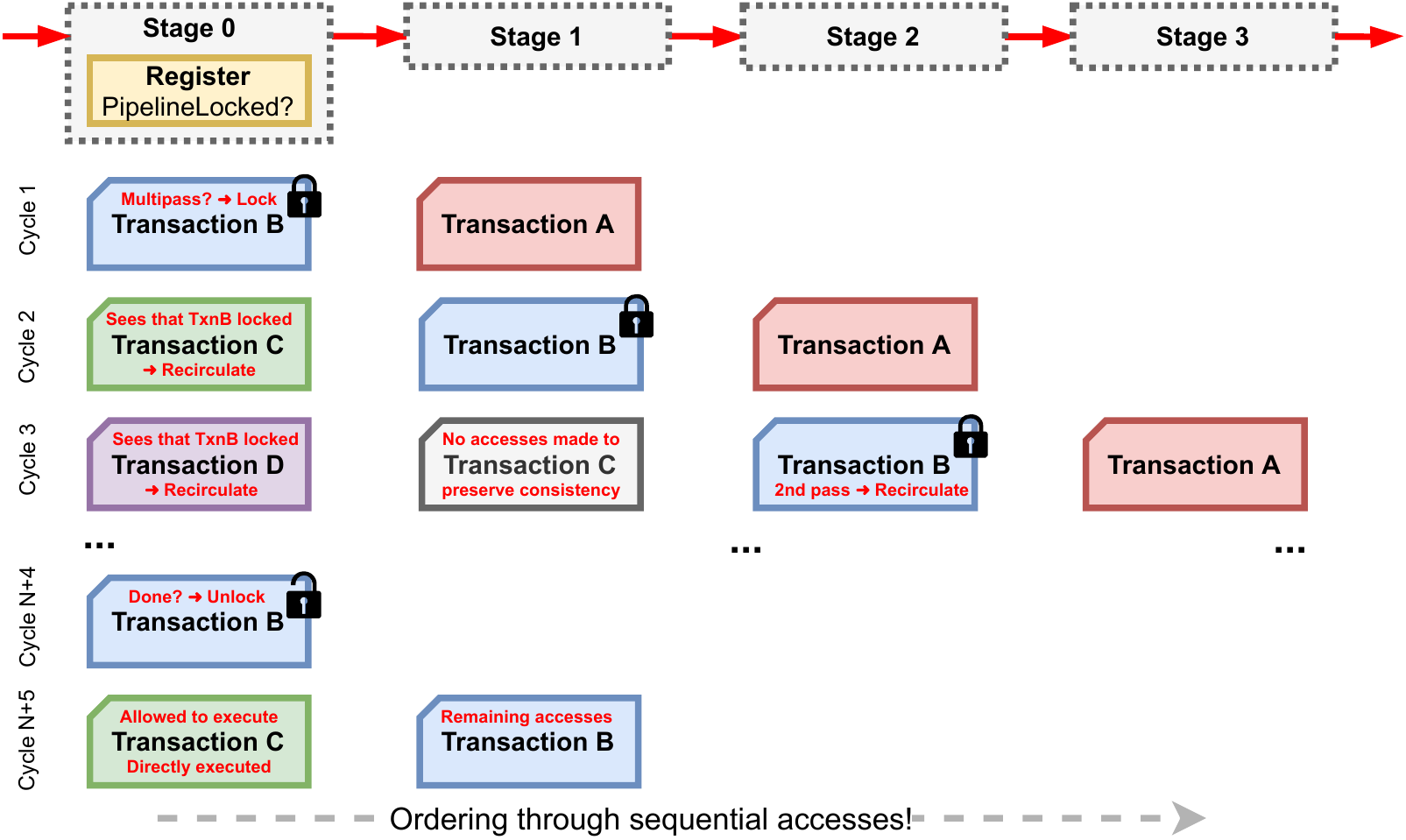}
    \caption{Processing of multi-pass transactions using a pipeline-lock. Each transaction checks the pipeline-lock in the first stage and can be recirculated to guarantee consistency of stored data, while one transaction is doing multiple passes through the switch pipeline.}
    \label{fig:switch_locking}
    \vspace{-4.5ex}
\end{figure}

\Cref{fig:switch_locking} shows an example with four transactions, where transaction $B$ is a multi-pass transaction and all other transactions are single-pass. The pipeline-state is shown for multiple cycles horizontally to show the flow of transactions over time. In cycle 1, the single-pass transaction $A$ is already in stage 1 when the multi-pass transaction $B$ arrives in stage 0. Hence, transaction $A$ is unaffected by the pipeline-lock that transaction $B$ acquires. However, the pipeline-lock of transaction $B$ then blocks the execution of following transactions $C$ and $D$ which are thus recirculated.
Once transaction $B$ starts its second pass though, it can directly unlock the pipeline-lock and thus $C$ and $D$ could be directly admitted to the pipeline.
This is possible because all transactions follow serial execution and can not overtake transaction $B$ in the pipeline. %

Again, this multi-pass scheme guarantees isolation and atomicity of transactions since the pipeline-lock enables a serial execution and prevents violation of conflicts.
Moreover, it also handles constraint checks to achieve consistency.
To execute constraint checks in this model, additional passes of a transaction are required (once the execution of the normal operations is complete) to not only check the constraints, but eventually also to  roll back changes in case a constraint is actually violated.

\subsection{Optimizations for Multi-Pass Transactions}
\label{sec:optimizations}

This section gives a short overview of optimizations we implemented for multi-pass transactions in \system{}. %

\vspace{-1.5ex}\paragraph{Fine-grained Locking} \label{sec:fine_locking} The switch's transaction throughput can be limited by the pipeline-lock that is used to preserve the consistency during the execution of multi-pass transactions.
One optimization is clearly to use more fine-grained locking (e.g., one lock per MAU stage) with multiple locks for different MAU stages.
With such a locking scheme, multiple transactions could run concurrently when they need locks for different disjunct registers in MAU stages.
However, while a single pipeline-lock can be implemented in P4 by using a single atomic stateful operation on a register, it is not trivial to implement multiple lock instances for one pipeline, because a packet can not access a register in a stage twice or undo lock-acquisition for all locks if for example one lock failed.

An obvious solution is to utilize a bit representation of the lock, e.g. 64 bits, where each bit represents a separate lock.
Unfortunately, Tofino's stateful register operations do not support bit-operations with variable masks as conditions for a compare-and-swap to test and set the lock.
Moreover, using a complex if-cascade to check each bit separately is also not supported since this would result in code that can not be executed in a single cycle on the switch (i.e., the P4-compiler fails to compile such programs).
However, with the current Tofino generation a 2-bit lock can be supported on the switch as shown in the pseudocode in \Cref{lst:twobitlock}. %

\begin{figure}
    \begin{minipage}{1.0\columnwidth} %
        \begin{lstlisting}[language=p4,caption={A 2-bit lock using a single register, in \pfour{}.},abovecaptionskip=-0.0ex,belowcaptionskip=0.0ex,label=lst:twobitlock,xrightmargin=.02\textwidth, xleftmargin=.02\textwidth]
struct lock_t { bit<8> left; bit<8> right; } // pair of locks
Register<...>(...) switch_lock;
RegisterAction<...>(switch_lock) try_lock = {
    void apply(inout lock_t value, out bit<1> rv) {
        if ((hdr.locks.left + value.left) == 2) {
            rv = 0;  // fail, left lock is already locked
        } else if ((hdr.locks.right + value.right) == 2) {
            rv = 0;  // fail, right lock is already locked
        } else {
            value.left  = value.left + hdr.locks.left;
            value.right  = value.right + hdr.locks.right;
            rv = 1;  // requested locks acquired successfully
        }
    }
};
\end{lstlisting}
        \vspace{-5.5ex}
    \end{minipage}
\end{figure}

\vspace{-1.5ex}\paragraph{Fast Recirculating} Programmable switches allow recirculating packets by virtually connecting the output of a port to the input of the same port. This recirculation port is used as a queue for multi-pass (or single-pass) transactions which need to wait for their execution. A main challenge of \system{} is to minimize the time a pipeline-lock is held.

Therefore, our design uses two recirculation ports. The first port is used exclusively for transactions which own the lock on the switch.
The second recirculation port is used for transactions which wait for execution. Since each port contains a separate queue for packets, transactions in the first recirculation port that hold a lock have a lower waiting time in the queue than the other (waiting) transactions, which reduces the time a lock is held by a transaction.

Finally, if a recirculation port receives more packets than it can handle, packets would be dropped. To prevent this, we actually split waiting transactions round-robin over multiple ports. To avoid starvation of recirculating transactions, each packet contains a recirculation counter, which can be used to prioritize the execution of transactions by the switch flow control as discussed before.

\vspace{-3.0ex}
\subsection{Processing Details}

In this section, we finally discuss several important details of how (single- and multi-pass) transactions are being  processed on the switch using network packets.
A network packet needs to contain all information that is necessary for executing the transaction on the switch. %
More precisely, switch packets in \system{} contain a header which holds important processing information followed by a variable number of instructions representing the operations of a transaction as shown in \Cref{fig:switch_instr}. These instructions are the operations that a transaction needs to run on the register arrays. Instructions can invoke trivial operations like read, writes or fixed point integer arithmetic as well as constrained-writes. %

In the following, we now discuss the fields which hold processing information individually (shown in gray in \Cref{fig:switch_instr}).
The first field \textit{is\_multipass} is a boolean flag which defines whether the incoming transaction can be executed in a single pass or not.
Following that, a \textit{locks} array shows the transaction engine which pipeline-locks need to be acquired in the first pass and freed in the last pass for multi-pass transactions.
For single-pass transactions, this field does not indicate which pipeline-locks need to be acquired but rather which pipeline-locks need to be free such that the transaction can be admitted to a pipeline.
The last field in the header is a counter \textit{nb\_recircs} to track the number of times a packet is recirculated.
It is incremented every time a transaction could not be executed and thus is recirculated. The switch flow control then uses this counter to prioritize long waiting transactions. %

Finally, it is important to note that the processing information in packets must be initialized by database nodes that issue a hot transaction. Therefore, we keep all relevant information directly on the database nodes.
For example, the information about the data layout (which tuple is stored in which MAU stage) is kept in an index structure redundantly per database node.
That way, a database node can decide whether a transaction can be executed in a single pass or if multiple passes are required.
More details about this are explained next in \Cref{sec:integration}.

\section{Integration with Host DBMS}
\label{sec:integration}

In this section, we discuss what is needed to integrate the switch execution into the host DBMS of \system{}.

\begin{figure}
    \captionsetup{aboveskip=0.0ex,belowskip=0.0ex}
    \centering
    \includegraphics[width=.95\columnwidth]{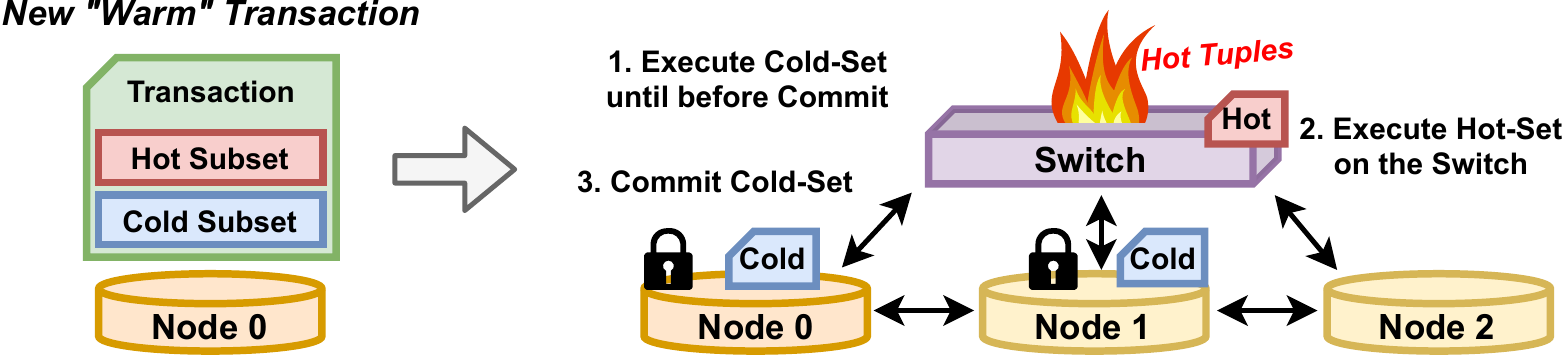}
    \caption{Processing of Warm Transactions. The switch transaction for the hot-set can only be sent when the operations on the cold tuples can not cause aborts anymore.}
    \label{fig:warm_txn}
    \vspace{-4.5ex}
\end{figure}

\subsection{Hot Transactions}
\label{sec:integration_hostextensions}
As a first aspect, we discuss how the execution of hot transactions can be integrated into the host DBMS. Warm transactions that span across cold and hot tuples will be discussed next.

\vspace*{-1.5ex}
\paragraph{Execution Scheme.} Most importantly, for executing hot transactions database nodes of the host DBMS need to trigger the execution of hot transactions. Hence, as a first extension, database nodes need to check for each transaction whether it accesses only tuples in the hot-set, otherwise the transactions is cold (or warm as we discuss next). 
To efficiently find out if a transaction accesses only hot tuples, we use an index that stores the keys of hot tuples. This index is replicated to all database nodes.
Since it is accessed frequently and its size is small, it resides in CPU caches most of the time.
Also, in the same index, we keep information in which stage and register a hot tuple is located on the switch.
This information is used for a hot transaction to determine if it is a single-pass transaction or not, and thus to set the header fields of the network packet accordingly. 
Finally, secondary indexes are supported by keeping them on the database nodes in \system{}; e.g., for a lookup the secondary key is first mapped to a primary key to find out if a hot tuple is accessed. Moreover, the secondary indexes are updated based on the results of the switch transaction (which is possible in our case since switch transactions cannot fail).

\vspace*{-1.5ex}\paragraph{Durability and Recovery.} 
\label{sec:durability}
A second important aspect is that the database nodes handle the durability and recovery for switch transactions.
To enable durability of switch transactions, database nodes use their local write-ahead log and append the operations for switch transactions they trigger to this log.
In addition, to enable a correct recovery of the switch state from the different local logs of database nodes (in case of a switch failure), the switch adds a unique transaction ID to each switch transaction that it executes, which represents the (serial) execution order of the transactions on the switch.
This ID is sent back together with the results of the read- and write-operations of a switch transaction to the database node in the response packet. The information is then appended to the write-ahead log of the database node.

In case the switch fails and needs to be recovered, the information in the local logs, which includes the globally ordered transaction IDs of the switch transactions, can then be used to recover a consistent state of the switch. 
A special case to consider for recovery are switch transactions that are in-flight (i.e., they are sent out but the result was not received by nodes). 
In \system{}, it is important to note that switch transactions count as committed before they are sent out since they cannot abort anymore. Therefore, a switch transaction and its intended read-/write-operations are appended to the log before the switch transaction is sent. The only info missing on database nodes at this point in the log is the unique transaction ID of switch transactions as well as the results of the read/write-operations.
To handle this case, \system{} aims to restore the order from dependencies in the read/write-set (as we discuss in more detail for warm transactions).
If no such dependency is detected, any order of switch transaction can be used during recovery for log replay.

\subsection{Warm Transactions}
\label{sec:integration_warmextensions}

\label{sec:warm_txns}

A second important aspect is how warm transactions are supported in \system{} that can span across both cold and hot tuples.

\vspace{-1.5ex}
\paragraph{Execution Scheme.} The main idea of the execution scheme for warm transactions is shown in shown in \Cref{fig:warm_txn}.
An important aspect of this scheme is the host DBMS first ensures that all operations on cold items will not abort before triggering the operations on the hot items on the switch using sub-transactions.

In \system{}, we use the following procedure to guarantee that operations on cold items cannot abort:
In the first step, \system{} acquires the locks on all cold items on the database nodes. %
Once all locks on cold items are acquired, \system{} then executes operations on the cold items and checks all constraints using a sub-transaction that runs on the database nodes only.
Once the sub-transaction on the cold items is ready to commit, \system{} sends out a network packet to the switch to trigger a sub-transaction, which then executes the operations on the hot items.
After receiving the executed switch transaction, the updates on the hosts on cold items are committed as we discuss below (see 2PC).

It is important to note that the execution model above does not support cases where warm transactions need to access cold tuples after warm tuples (e.g., 
a write to a cold tuple that depends on a read of a hot tuple).
We support this case in \system{} by additionally offloading those cold tuples to the switch during the offload phase of \system{}.
For the cold tuples stored on the switch, we then use the very same processing scheme as for hot tuples (i.e., they are treated as if they were hot).

\begin{figure}
    \captionsetup{aboveskip=0.0ex,belowskip=0.0ex}
    \centering
    \includegraphics[width=.99\columnwidth, trim=0mm 0mm 0mm 0mm, clip]{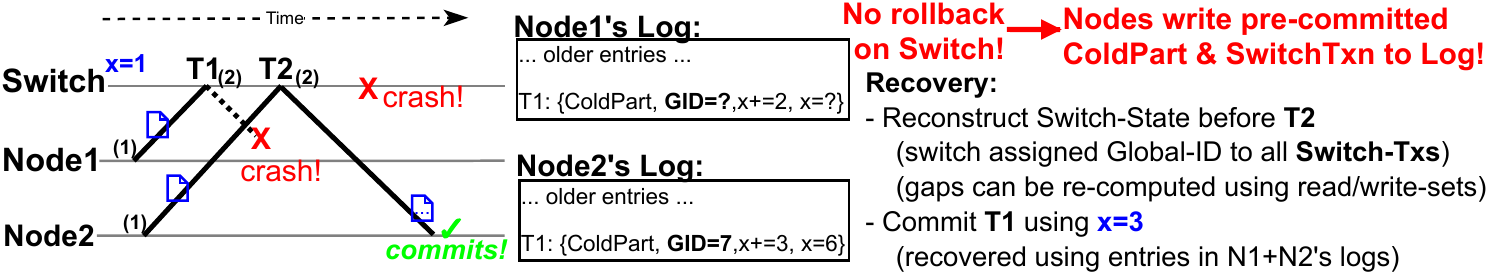}
    \caption{Durability and Recovery of Warm Transactions. In this scenario, two warm transactions (T1 and T2) are executed. During execution, Node1 and the Switch both crash.} %
    \label{fig:durability_recovery_1}
    \vspace{-3.5ex}
\end{figure}

\vspace{-1.5ex}\paragraph{Durability and Recovery.}
Durability and recovery for warm transactions is handled similarly as for hot transactions. 
The only difference is that we need to consider additional failure cases. 
In the following, we discuss the most complex case where both a node and the switch that are involved in a warm transaction fail after switch transactions were sent out and before the result is received. %
\Cref{fig:durability_recovery_1} shows an example for this case.
In this scenario, the globally-unique transaction ID of T1 is missing in the log of Node1 and therefore the execution order of T1 and T2 cannot be restored during recovery, because it is not clear if T1 was executed before or after T2.
However, we can still reconstruct the order of the two switch transactions for recovery by analyzing the logs of all database nodes for dependencies using the read/write-sets. For example, as shown in \Cref{fig:durability_recovery_1}, the order (T1 before T2) could be restored from the read/write-set of T2 in the log of Node2 which indicates that T2 reads $x{=}3$ that must have been written by T1.
If there were no dependency between T1 and T2, any order could be used \footnotemark[1]. %

\footnotetext[1]{Additional failure cases, are discussed in the appendix.}

\begin{figure}
    \captionsetup{aboveskip=0.0ex,belowskip=0.0ex}
    \centering
    \includegraphics[width=.92\columnwidth, trim=0mm 0mm 0mm 0mm, clip]{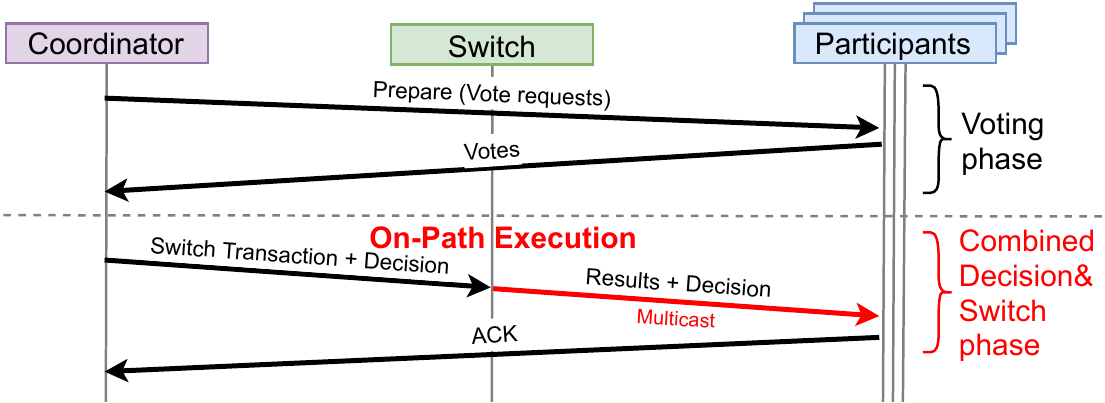}
    \caption{Coordination of the distributed commit of warm transactions. Once the coordinator knows that the nodes (i.e., participants) can commit, it sends out the switch-transaction. After execution, the switch sends the commit decision via a multicast to the nodes.
    }
    \label{fig:2pc_with_switch}
    \vspace{-4.5ex}
\end{figure}

\vspace{-1.5ex}\paragraph{Integration with 2PC}
\label{sec:two_phase_commit}
Finally, a last important aspect that is different for warm transactions compared to hot transactions is that we need to coordinate the commit across multiple nodes (in case the cold part is distributed) and the switch.
To coordinate the distributed commit of warm transactions, we extend the classical Two-Phase-Commit (2PC) protocol in \system{} as shown in \Cref{fig:2pc_with_switch}.
Similar to the traditional 2PC, in our scheme a coordinator decides whether a warm transaction can commit based on the votes of all participating nodes similar to the normal 2PC. However, in addition if all nodes voted on commit, the switch sub-transaction is sent out by the coordinating node to the switch. 
The switch then executes the switch transaction and broadcasts the results using a multicast operation to all database nodes, which then finally commit their results without requiring additional round-trips.
In case only one database node is involved in a warm transaction, we do not rely on 2PC. Instead, the switch sub-transaction can be sent out by the coordinator without executing a voting phase.

An interesting point of our integration with the 2PC protocol is that sub-transactions on the switch can directly commit once they are executed; i.e., sub-transactions on the switch do not need to wait for the commit on the database nodes.
Note that this is not a problem since switch sub-transactions of warm transactions will commit in the same order as the host sub-transactions of warm transactions. 
The reason is that in case there is a conflict on cold-items between two warm transactions, only the warm transaction which holds the locks on the conflicting cold items will send out the sub-transaction on the hot items.
Otherwise, if there is no conflict of warm transactions on the cold tuples, then the sub-transactions on the hot-items can be executed and committed independently without coordination.

\section{Experimental Evaluation}
\label{sec:evaluation}

In the following, we present the results of our evaluation of \system{}. We first compare \system{} to other baselines using three OLTP benchmarks (YCSB, SmallBank, TPC-C).
Afterwards, we then present microbenchmarks that show a more in-depth analysis of \system{}.

\subsection{Setup and Baselines}
\label{sec:setup}
\paragraph{Setup} The experimental setup consists of an 8-node cluster, where 6 of the nodes are equipped with two Intel(R) Xeon(R) Gold 5120 2.2GHz CPUs and 2 nodes are equipped with two Intel(R) Xeon(R) Gold 5220 2.2GHz CPUs.
Each node is connected to an Intel Tofino switch (BF2556X-1T) \cite{bf2556x} via a 10G NIC (Intel X540-AT2). Hyper-threading is enabled for all nodes, but only one socket with 256GB of main-memory is used to avoid cross-socket communication to the NIC.
The operating system is Ubuntu 18.04.1 LTS with Linux kernel 4.15.0 installed on all nodes. \system{} is implemented in C++20 and compiled with gcc-10. We utilize DPDK 20.08 \cite{dpdk} for optimized network performance and communication between the database nodes.
The switch's control-plane logic of \system{} is implemented in C++ and the switch's data plane logic of \system{} is implemented in \pfour{} and compiled using Intel SDE-9.3.0 \cite{bfsde}.

\vspace{-1.5ex}\paragraph{\system{} Implementation} Our prototype implementation of \system{}\footnote[2]{\url{https://github.com/DataManagementLab/p4db}} includes the switch that we integrated into a shared-nothing distributed in-memory DBMS as discussed in \Cref{sec:integration}. To allow the execution of cold transactions in \system{}, we implemented two commonly used variants of Two-Phase Locking (2PL) with deadlock prevention: \textit{NO\_WAIT} aborts a transaction as soon as a lock request is denied. For \textit{WAIT\_DIE}, transactions are assigned a unique timestamp at transaction start and transactions only wait for a lock if it is owned by a transaction with a younger timestamp, otherwise it aborts. %
Similar to other recent papers \cite{appuswamy2017analyzing,harding2017evaluation,zamanian2020chiller}, we do not include other deadlock detection algorithms such as \textit{DL\_DETECT} since these schemes are more complex and typically do not provide any significant benefits.

\vspace{-1.5ex}\paragraph{Baselines}\label{sec:baseline} As a first baseline, we use the same prototype, however, our switch acts only as a traditional network switch and thus does not execute in-switch transactions (and hence the baseline is called \emph{No-Switch}).
As a second baseline, we used another variant of our prototype and use the switch only as a lock manager for hot tuples as suggested in \cite{yu2020netlock} (called \emph{LM-Switch}).

\subsection{Workloads}
\label{sec:workloads}
For our experiments, we used three different OLTP benchmarks (YCSB, SmallBank, TPC-C).
As the performance metric, we report throughput of committed (hot and cold) transactions per second while we also discuss latency at the end of the experiments.
In the following, we describe each workload and how we defined the hot-set of tuples based on prior papers.
Moreover, for all benchmarks, we vary the transactional load (i.e., the number of worker threads per database node) to analyze the effect of increased contention on the hot-set.
In particular, we use configurations from $8$ worker threads (normal load) to $20$ worker threads (high load) per node, each executing a single transaction at a time.

\vspace{-1.5ex}\paragraph{YCSB}
The Yahoo! Cloud Serving Benchmark \cite{cooper2010benchmarking} consists of a single table that is partitioned round-robin across all nodes.
Similar to other papers \cite{yu2014staring,zamanian2019rethinking}, we define a transaction as a group of 8 read/write-operations.
We also populate the table with 1 billion entries consisting of an 8 byte key and 8 byte value.

In our evaluation, we use workloads A-C which are characterized by different ratios of read/write operations within a transaction to mimic update-heavy or read-heavy workloads (A - Update heavy 50/50, B - Read heavy 95/5, C - Read only 100/0 ). %
To simulate skew in the access patterns of tuples, similar to \cite{yu2018sundial}, we use a hot-set consisting of 50 key-value pairs per database node which receive 75\% of all accesses.

\vspace{-1.5ex}\paragraph{SmallBank}
The SmallBank benchmark models a banking application where transactions perform different operations on 1 or 2 customer accounts. The original workload contains 5 different types of transactions \cite{alomari2008cost}.
Similar to \cite{difallah2013oltp,kallman2008h}, we additionally include a \textit{Payment} transaction that transfers money between two accounts. %

Compared to YCSB, SmallBank represents a workload with a fixed read-ratio of 15\% and uses 1 million bank accounts. However, the difference to YCSB is that it contains read-dependent-writes (as well as some simple constraints), which makes it more complex to implement and shows the need for the declustered data layout in \system{}. %
To simulate skew in the access patterns of tuples, similar to \cite{kalia2016fasst}, we used different hot-sets that vary in their sizes between $5$ and $15$ tuples per database node and that are accessed by 90\% of all transactions.

\vspace{-1.5ex}\paragraph{TPC-C}
The TPC-C benchmark \cite{tpcc} is the most complex benchmark in our evaluation and consists
of 9 tables and 5 transaction types.  %
Although TPC-C is designed to be highly partitionable by warehouse, it still contains several contention points. For example, the \textit{NewOrder} transaction increments the global \textit{next-order-id} per district and the \textit{Payment} transaction updates the total balance per warehouse that causes contention on individual tables.

For our evaluation, we use a mix of only \textit{NewOrder} and \textit{Payment} transactions, similar to other papers \cite{bang2020tale,zamanian2020chiller,yu2014staring}, since these account for $90\%$ of the transactional workload.
Due to their complexity and the mix of cold and hot tuples within transactions, the logic of this workload requires warm transactions, unlike the previous benchmarks, and thus shows a different aspect of \system{} than the benchmarks before.

\subsection{YCSB Experiments}
\label{sec:ycsb}

In this experiment, we first validate our approach with YCSB as the most comprehensible OLTP benchmark. More precisely, we compare the throughput for YCSB using \system{} with its in-switch execution of hot transactions with the other two baselines; i.e., switch as lock-manager (\emph{LM-Switch}) and a traditional distributed DBMS without using the switch for OLTP (\emph{No-Switch}).

\begin{figure}
    \captionsetup[subfigure]{aboveskip=0.0ex,belowskip=0.0ex}
    \captionsetup{aboveskip=0.0ex,belowskip=0.0ex}
    \begin{subfigure}{.95\columnwidth}
        \centering
        \includegraphics[width=0.95\linewidth, trim=0mm 0mm 0mm 0mm, clip]{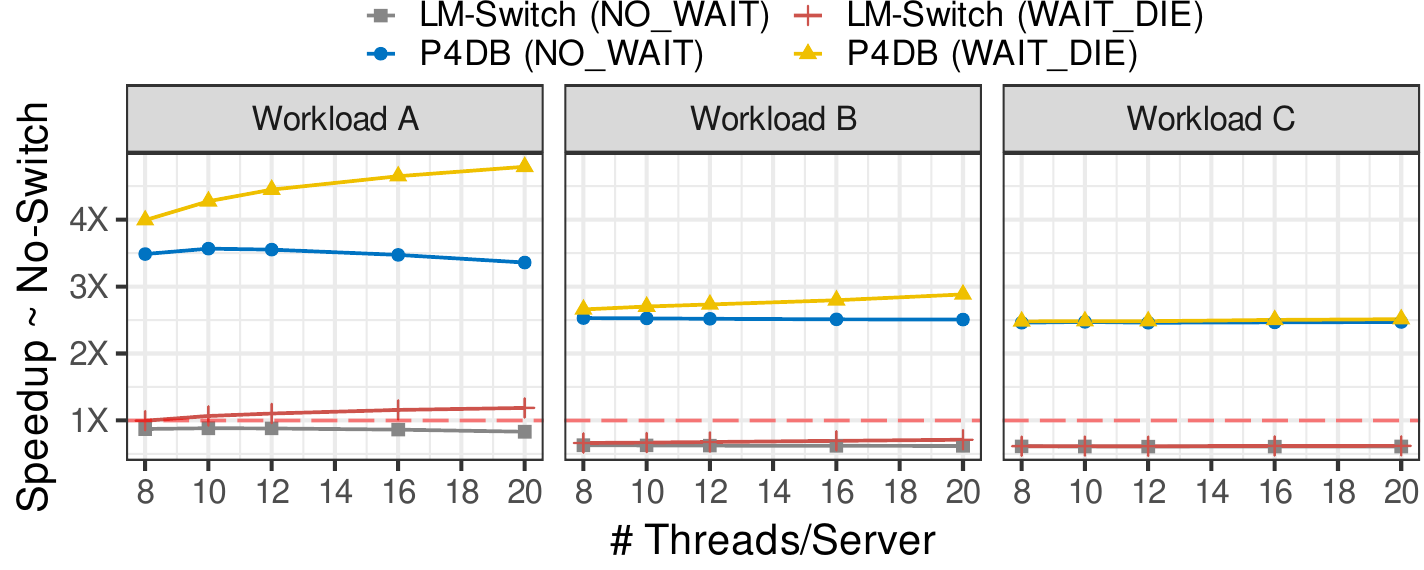}
    \end{subfigure}
    \begin{subfigure}{.95\columnwidth}
        \centering
        \includegraphics[width=0.95\linewidth, trim=0mm 0mm 0mm 8mm, clip]{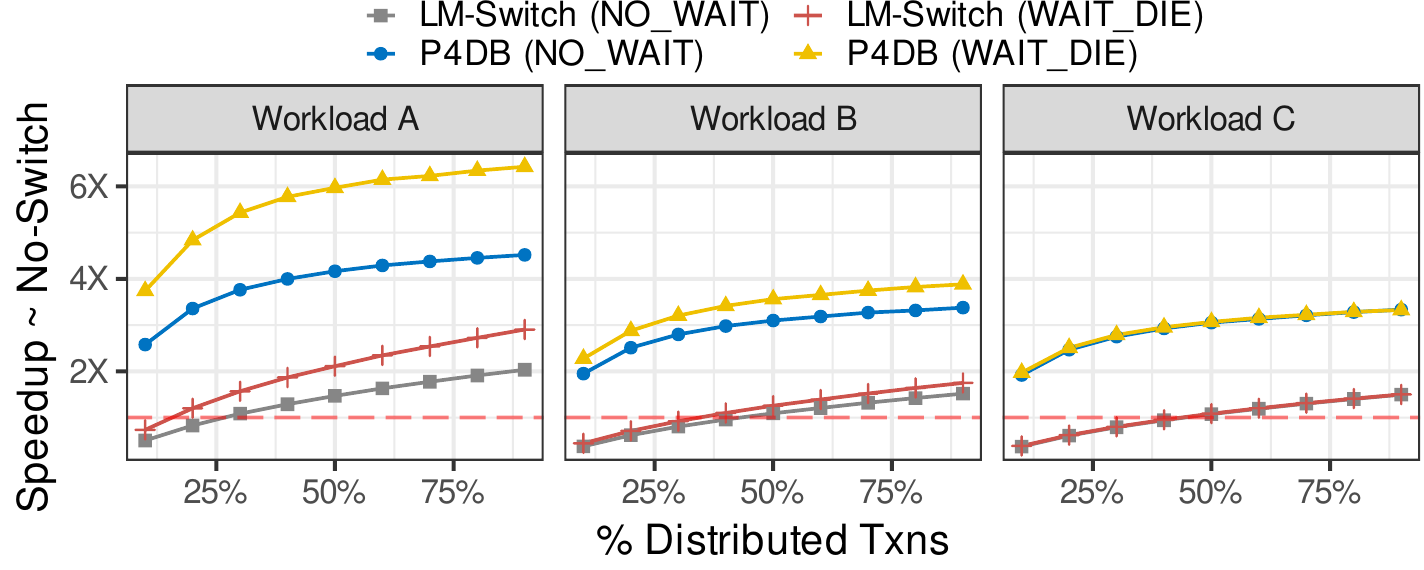}
    \end{subfigure}
    \caption{YCSB Results --- The speedup of \system{} and \emph{LM-Switch} over \emph{No-Switch}. \system{} yields the best improvements for increased contention (upper row) and an increased ratio of distributed transactions (lower row).}
    \label{fig:ycsb}
    \vspace{-2.5ex}
\end{figure}

\vspace{-1.5ex}\paragraph{Varying Contention}
\label{sec:eval_switch_lm}
In the first experiment, we compare the performance of \system{} and the baselines for a varying degree of contention.
To increase the contention, we scale the number of worker threads as described before.
\Cref{fig:ycsb} (upper row) compares the speedup of \system{} w.r.t throughput over the \emph{No-Switch} baseline, as well as the speedup of the \emph{LM-Switch} over the same baseline.
As we can see, \system{} outperforms both baselines and provides a speedup of up to $5\times$ over the \emph{No-Switch} baseline for the highest contention cases.
The speedup of \system{} is the highest for workload A, due to its higher write-ratio causing decreased performance of both baselines.

Different from \system{}, the baseline which uses the switch as lock-manager (\emph{LM-Switch}), does not yield any significant speed-up.
In general, while \emph{LM-Switch} enables high throughput for uniform workloads as shown in \cite{yu2020netlock}, this approach provides only minimal benefit for skewed workloads.
The reason is that the overall latency of how long locks are held on contended items is not really reduced with this approach, which is very different from the coordination-free model of \system{}.

\vspace{-1.5ex}\paragraph{Varying Distributed Transactions}
Another aspect that is important in the context of \system{} is distributed transactions, since they could further amplify the effect of contention (due to higher access latencies of hot tuples). To investigate further the results from the previous experiment and to show the benefits of our in-switch approach for distributed transactions, we vary the degree of distributed transactions using the maximum number of worker threads from the previous experiment per node. %

The results of this experiment can be seen in \Cref{fig:ycsb} (lower row) which shows the speedup for \system{} and \textit{LM-Switch} over \textit{No-Switch}. Again, as before, \system{} provides a significant speedup while \textit{LM-Switch} provides only limited gains for similar reasons.
An interesting observation is that a higher degree of distributed transactions results in an increased throughput in \system{} leading to a speedup of more than $6\times$ for 100\% distributed transactions due to the lower network latency (i.e., only ½ RTT) for hot tuples.
Furthermore, at 0\% distributed transactions, \system{} still provides a significant speedup over the \textit{No-Switch} baseline.
Here, the main reason is that the switch is much more efficient in processing transactions on the hot items than a normal database node.

\begin{figure}
    \captionsetup[subfigure]{aboveskip=0.0ex,belowskip=0.0ex}
    \captionsetup{aboveskip=0.0ex,belowskip=0.0ex}
    \centering
    \includegraphics[width=.95\columnwidth, trim=0mm 0mm 0mm 0mm, clip]{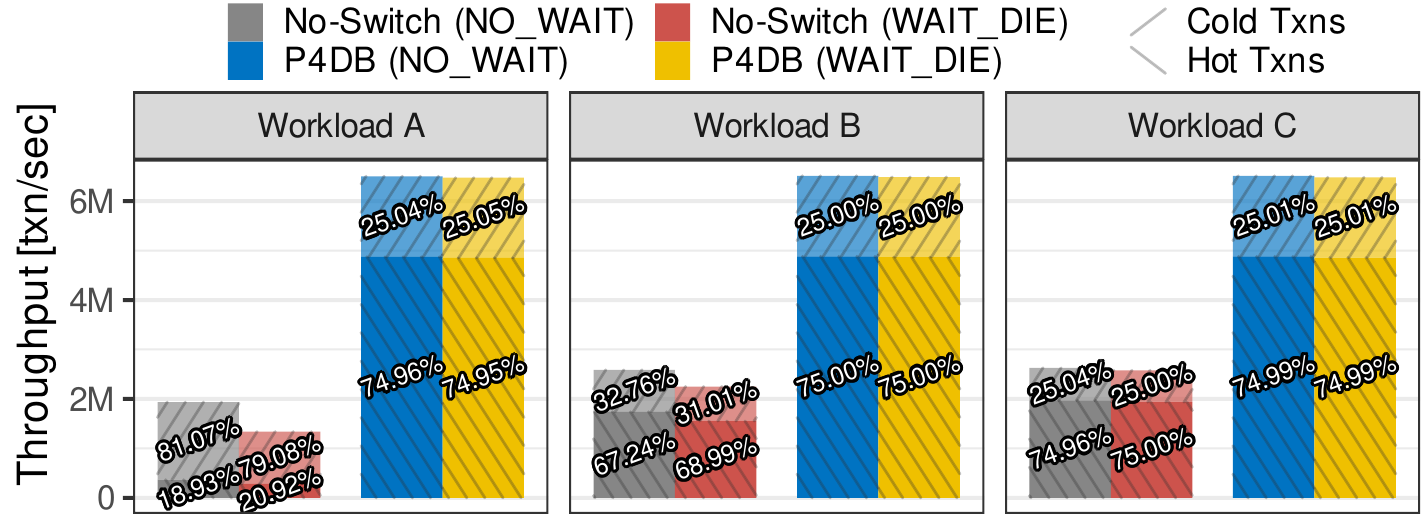}
    \caption{Benefits of switch transactions in YCSB. The switch executes the 75\% hot transactions in the workload. In a traditional DBMS without switch a majority of those is aborted due to contention.}
    \label{fig:ycsb_comparison}
    \vspace{-3.5ex}
\end{figure}

\vspace{-1.5ex}\paragraph{Break-down of Hot and Cold Tx's}
To better understand the effects of why \system{} dominates the throughput, we now present a throughput break-down of committed hot/cold transactions.
For this experiment, we analyze the throughput (absolute results) for 20\% distributed transactions and $20$ worker threads per database node, as in the previous experiment.
We decided to not show the \textit{LM-Switch} baseline, since the break-down is very similar to the one of \textit{No-Switch}.

\Cref{fig:ycsb_comparison} shows the results of this experiment, where the fractions are marked for each bar. We see that for the update-heavy workload A, with \textit{No-Switch} only a low fraction of hot transactions commits (dark grey, dark red) due to contention. \system{}, on the other hand, has a much higher commit ratio for hot transactions of around 75\%  (dark blue, dark yellow).
For workload C, which is a pure read-only workload, the ratios for hot transactions in \textit{No-Switch} and \system{} are the same (at approx. $75\%$) since no aborts happen for \textit{No-Switch} on hot tuples.
Nevertheless, \system{}'s overall throughput is higher in all workloads, including workload C.
Moreover, the switch throughput in \system{} is not affected by different read-write ratios of the workloads $A$-$C$, since all transactions can be executed (lock-free) in a single pass.

\subsection{SmallBank Experiments}

In this experiment, we next compare the throughput for SmallBank using \system{} to \emph{No-Switch}, which, as mentioned earlier, represents a more challenging workload as YCSB due to read-dependent-writes that could lead to multi-pass transactions.
For executing this workload in \system{}, we use an optimal declustered data layout. We also show the effect of a non-optimal data layout in our microbenchmarks later.
Furthermore, we dropped the \textit{LM-switch} baseline from all subsequent experiments since it similarly shows only minimal gains over \emph{No-Switch} as for YCSB.

\vspace{-1.5ex}\paragraph{Varying Contention}
Similar as for YCSB, we first start with an experiment where we increase the load and thus the contention on a fixed set of hot tuples. %
Additionally, we use different hot-set sizes (5, 10, 15 hot tuples per database node) instead of different read/write ratios that we evaluated for YCSB.

\begin{figure}
    \captionsetup[subfigure]{aboveskip=0.0ex,belowskip=0.0ex}
    \captionsetup{aboveskip=0.0ex,belowskip=0.0ex}
    \begin{subfigure}{.95\columnwidth}
        \centering
        \includegraphics[width=.95\columnwidth, trim=0mm 0mm 0mm 0mm, clip]{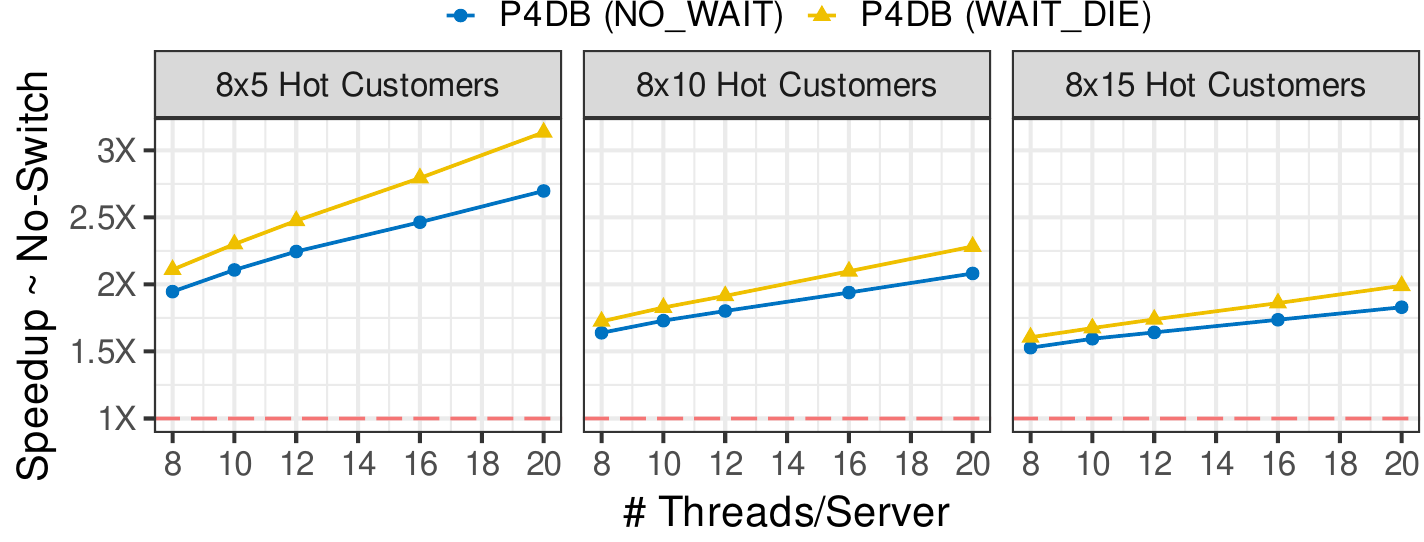}
        \phantomsubcaption
        \label{fig:smallbank_scaleout}
    \end{subfigure}
    \begin{subfigure}{.95\columnwidth}
        \centering
        \includegraphics[width=0.95\linewidth, trim=0mm 0mm 0mm 5mm, clip]{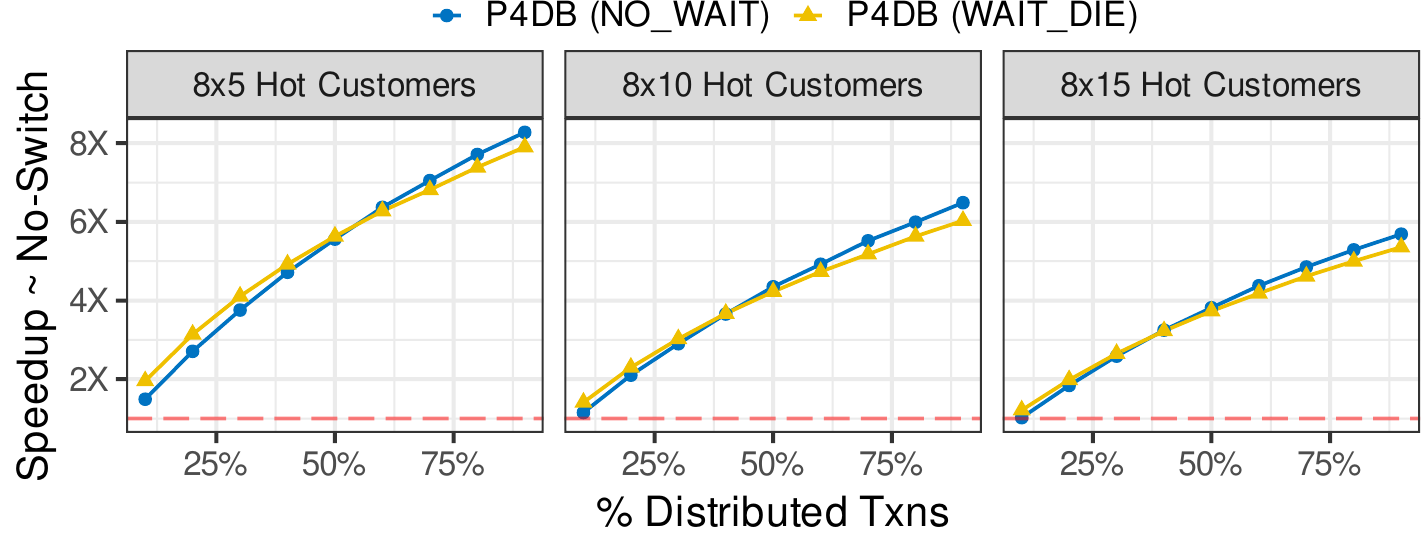}
        \phantomsubcaption
        \label{fig:smallbank_distributed}
    \end{subfigure}
    \caption{SmallBank Results --- \system{} achieves a near-linear speedup for increased contention (upper row) and an increased ratio of distributed transactions (lower row).}
    \label{fig:smallbank}
    \vspace{-2.5ex}
\end{figure}

The results of this experiment are shown in \Cref{fig:smallbank} (upper row).
In case for the smallest hot-set, \system{} again provides a speedup of up to 3$\times$ for the largest number of worker threads over the \textit{No-Switch} baseline. Especially for the smallest hot-set, \system{} achieves high switch-throughput due to its single-pass execution model.
Moreover, as the hot-set size increases and the contention decreases, \system{} still can provide a significant speedup of up to $2\times$.

\vspace{-1.5ex}\paragraph{Varying Distributed Transactions}
Analogous to YCSB, we vary the amount of distributed transactions for SmallBank while fixing the number of workers to $20$ per database node. %
\Cref{fig:smallbank} (lower row) shows the results.
With increasing fraction of remote transactions, we again see a speedup of \system{} over \textit{No-Switch} even though the workload is more challenging as YCSB as discussed before due to read-dependent-writes.
However, due to our declustered data layout of \system{} on the switch, all hot transactions in SmallBank can be executed efficiently in a single pass on the switch, resulting in a similar speedup as before for YCSB.

\subsection{TPC-C Experiments}
\label{sec:eval_tpcc}

In this experiment, we use TPC-C as a complex benchmark.
As mentioned before, in TPC-C transactions cannot be implemented as solely hot or cold transactions, but it requires warm transactions that span across hot and cold tuples to include the switch in the transaction processing. To be more precise, we offloaded all contended columns of the warehouse and district tables with write-accesses as well as stock columns of most ordered items to the switch.
An interesting question thus is how much gains \system{} provides given this more challenging workload.

\begin{figure}
    \captionsetup[subfigure]{aboveskip=0.0ex,belowskip=0.0ex}
    \captionsetup{aboveskip=0.0ex,belowskip=0.0ex}
    \begin{subfigure}{.95\columnwidth}
        \centering
        \includegraphics[width=0.95\linewidth, trim=0mm 0mm 0mm 0mm, clip]{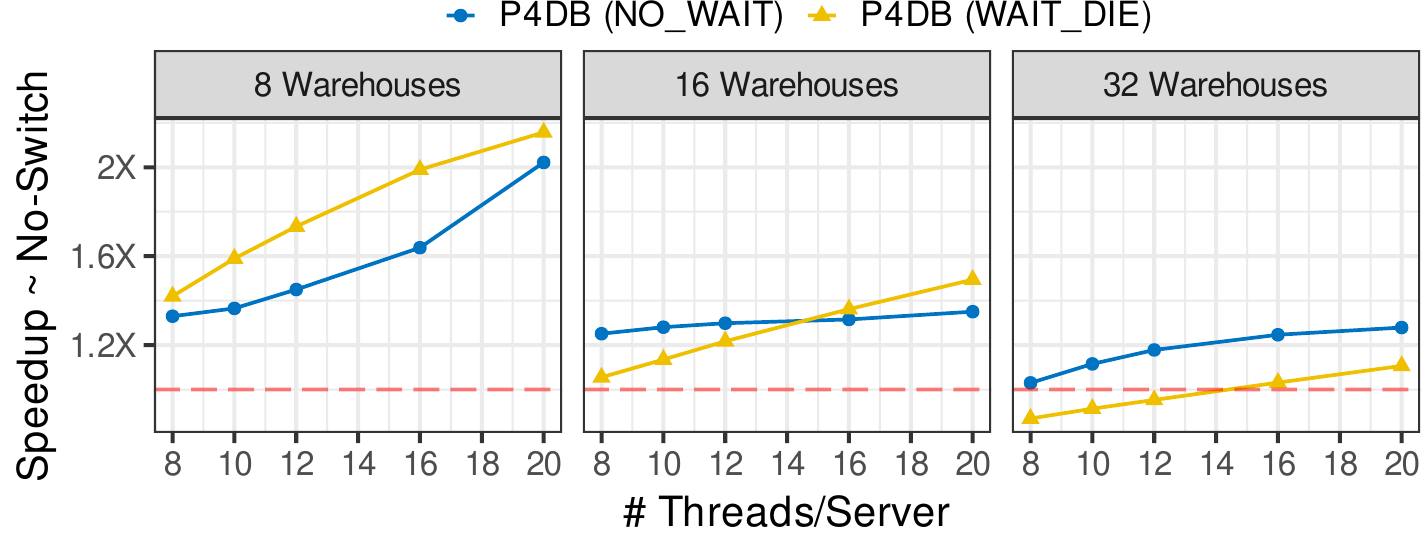}
        \phantomsubcaption
        \label{fig:tpcc_scaleout}
    \end{subfigure}
    \begin{subfigure}{.95\columnwidth}
        \centering
        \includegraphics[width=0.95\linewidth, trim=0mm 0mm 0mm 5mm, clip]{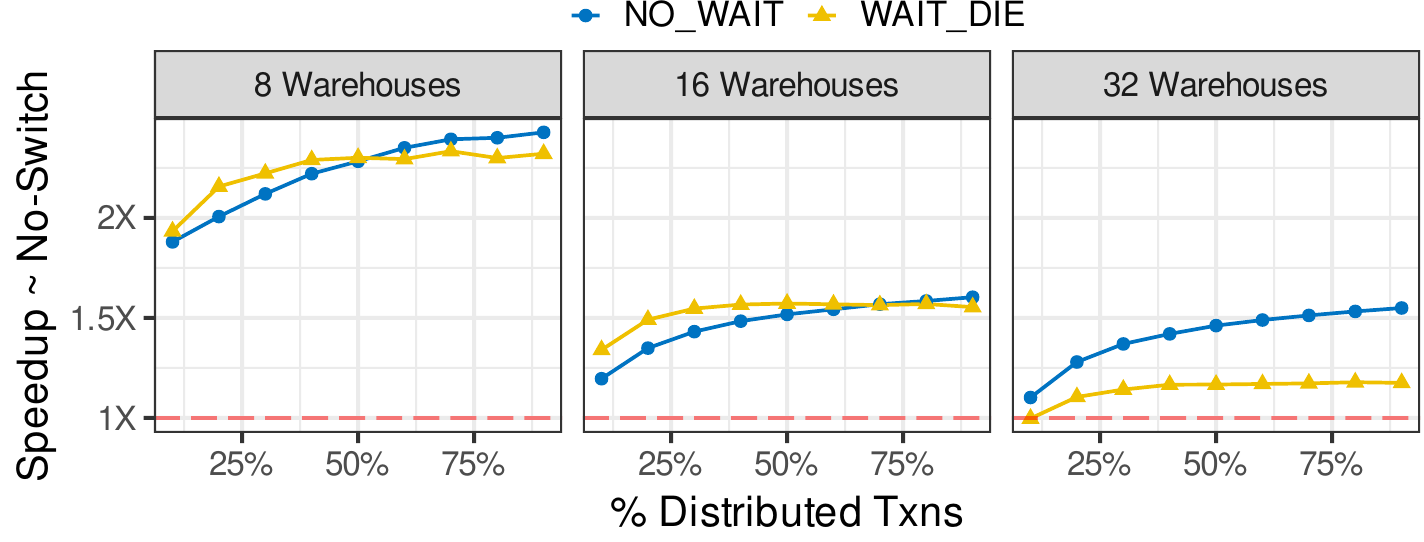}
        \phantomsubcaption
        \label{fig:tpcc_distributed}
    \end{subfigure}
    \caption{TPC-C Results --- Even though warm transactions are used, \system{} yields a significant speedup for increased contention (upper row) and an increased ratio of distributed transactions (lower row).}
    \label{fig:tpcc}
    \vspace{-3.5ex}
\end{figure}

\vspace{-1.5ex}\paragraph{Varying Contention}
The goal with TPC-C is to show whether using the switch as an accelerator when using warm transactions helps to increase overall system throughput.
Again, we first scale the amount of worker threads per database node to show the effect of an increased contention while using a fixed ratio of $20\%$ distributed transactions.
We additionally vary the number of warehouses as they influence contention as well.

The results of this experiment are shown in \Cref{fig:tpcc} (upper row).
Overall, we see that with the highest contention configuration of 8 warehouses and 20 worker threads, \system{} still achieves a significant speedup of more than $2\times$.
However, clearly, when increasing the number of warehouses to 16 and 32, the contention gets lower which decreases the gains.
For example, with 32 warehouses, the speedup decreases to $1.3\times$.

\vspace{-1.5ex}\paragraph{Varying Distributed Transactions}
This experiment shows the impact of a varying degree of distributed transactions when using warm transactions.
For \textit{NewOrder}, we varied the probability that an ordered item is located in a remote warehouse and for \textit{Payment} that the paying customer is located at a remote warehouse.

\Cref{fig:tpcc} (lower row) shows the results. Again, as for YCSB and SmallBank when increasing the ratio of distributed transactions, the access conflicts for \textit{No-Switch} become more visible since more of the warm transactions need to wait for network round-trips and thus the latency for those transactions increases. In contrast, \system{} reduces the network round-trip cost for hot items, which leads to speedup increases of up to $2.5\times$ compared to \textit{No-Switch}.

\subsection{Microbenchmarks}

This section presents microbenchmarks to further understand details of the transaction execution on the switch when using \system{}.

\vspace{-1.5ex}\subsubsection{Varying Hot/Cold Ratio}
\label{sec:eval_skew}

In the previous experiments, we used a fixed ratio of hot to cold transactions.
Thus in this microbenchmark, we further investigate the impact using different ratios. As a workload, we show the results for YCSB-A with 20\% remote transactions, but the findings generalize to other workloads. %

\Cref{fig:skew_throughput} shows the absolute throughput, while \Cref{fig:skew_speedup} shows the corresponding speedup of \system{} over \textit{No-Switch}. As we see, when the fraction of hot transactions increases, the increased contention on hot tuples leads to a lower throughput of \textit{No-Switch} caused by a higher abort rate. In stark contrast, \system{}'s throughput increases with more hot transactions, resulting in a speedup of more than $50\times$ for $100\%$ hot transactions.
With $0\%$ hot transactions, \system{} and \textit{No-Switch} have a similar performance because the switch then only redirects cold transactions to database nodes and does not process any hot transactions.

\vspace{-1.5ex}\subsubsection{Optimizations of Switch Processing}
\label{sec:eval_optimizations}

In this microbenchmark, we study the effect of the different optimizations we outlined in \Cref{sec:optimizations}.
For evaluating these effects in isolation, we use the hot (i.e., switch-only) transactions of the YCSB-A workload only.
As a baseline (\textit{Unoptimized}), we use a random data layout without any of the optimizations in \Cref{sec:optimizations} (i.e., fast recirculation and fine-grained locking turned off).
We then turn on these optimizations one at a time, which optimize the efficiency of multi-pass transactions.
Finally, we apply our optimal data layout, which aims to avoid multi-pass transactions as much as possible.

\begin{figure}
    \captionsetup[subfigure]{aboveskip=0.0ex,belowskip=0.0ex}
    \captionsetup{aboveskip=0.0ex,belowskip=0.0ex}
    \begin{subfigure}[b]{.32\columnwidth}
        \centering
        \includegraphics[width=0.99\linewidth, trim=0mm 0mm 0mm 0mm, clip]{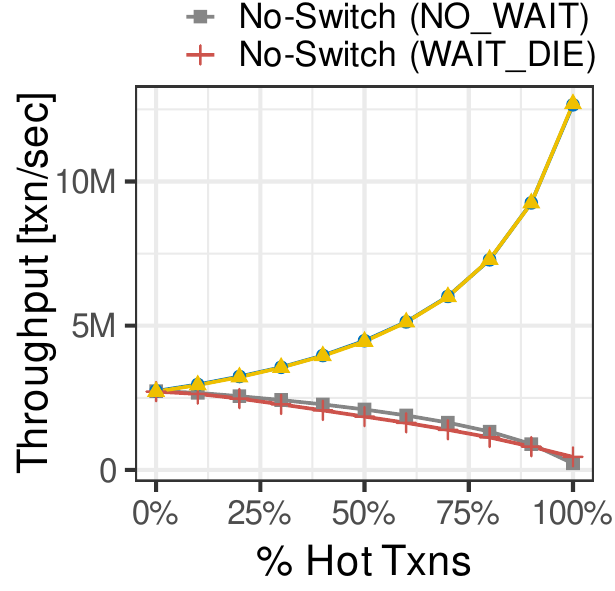}
        \caption{Skew: Throughput}
        \label{fig:skew_throughput}
    \end{subfigure}
    \begin{subfigure}[b]{.32\columnwidth}
        \centering
        \includegraphics[width=0.99\linewidth, trim=0mm 0mm 0mm 0mm, clip]{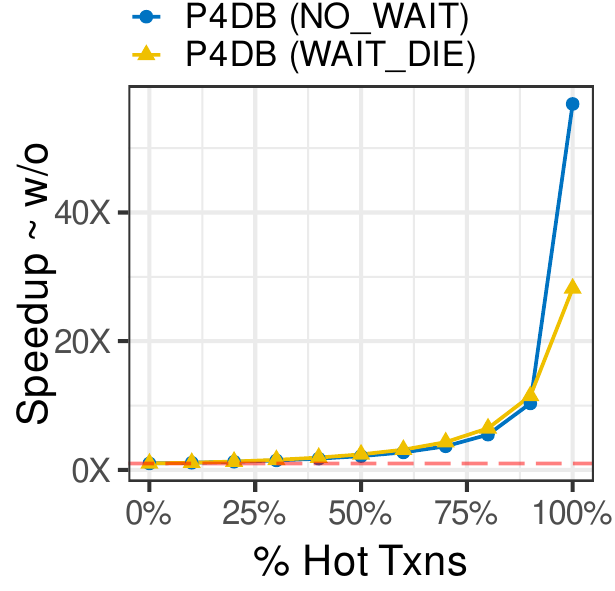}
        \caption{Skew: Speedup}
        \label{fig:skew_speedup}
    \end{subfigure}
    \begin{subfigure}[b]{.30\columnwidth}
        \centering
        \includegraphics[width=0.99\linewidth, trim=0mm 0mm 0mm 0mm, clip]{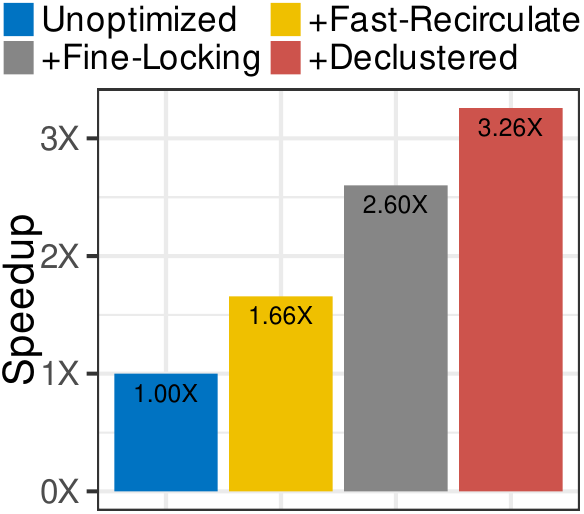}
        \vspace{0.0ex}
        \caption{Optimizations}
        \label{fig:optimizations}
    \end{subfigure}
    \caption{Microbenchmarks --- (a-b) shows the effect of a higher \% of hot transactions and (c) shows the impact of optimizations for multi-pass transactions within \system{}.}
    \label{fig:microbench}
    \vspace{-2.5ex}
\end{figure}

The results are shown in \Cref{fig:optimizations}. As we can see, the first optimization called \emph{Fast-Recirculate}, which creates a fast path for recirculating transactions that already own a pipeline-lock, leads to a speedup of approximately $1.66\times$ in \system{}.
As a second optimization, we then applied fine-grained locking, which allows that two multi-pass transactions can be executed at the same time.
This optimization together with the previous one results in a speedup of $2.6\times$.
Finally, when applying our optimal data layout, we can further boost the performance since the need to use multi-pass transactions is significantly reduced. As a result, when turning on all optimizations and using the optimal data layout, we see an overall speedup of $3.26\times$ compared to the baseline (\textit{No Optimization}).

\vspace{-1.5ex}\subsubsection{Single- vs. Multi-pass switch transactions}
\label{sec:eval_partitioning}

\begin{figure}
    \captionsetup[subfigure]{aboveskip=0.0ex,belowskip=0.0ex}
    \captionsetup{aboveskip=0.0ex,belowskip=0.0ex}
    \centering
    \includegraphics[width=0.99\columnwidth]{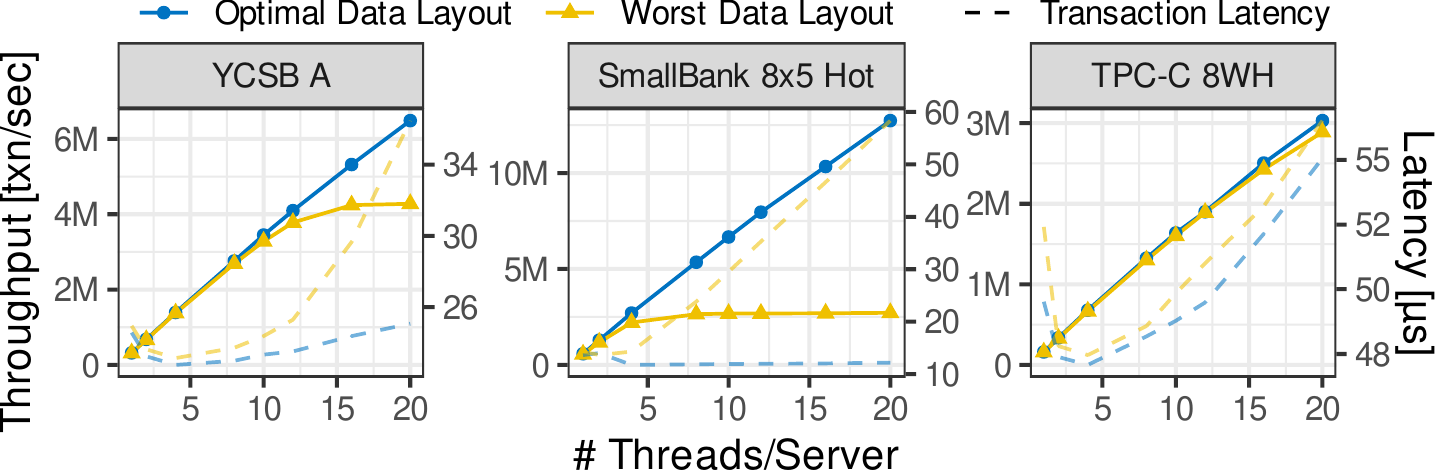}
    \caption{Impact of the optimal data layout for the three workloads. SmallBank shows the highest benefits, while TPC-C is limited by warm transactions and hence the data layout has almost no effect. }
    \label{fig:exp_partitioning}
    \vspace{-3.5ex}
\end{figure}

As shown before, our data layout efficiently helps to improve the throughput of \system{} by enabling single-pass execution for more transactions.
While in the previous experiment we showed the effect only for the YCSB-A workload, in this experiment we use all three workloads to compare the effect of our data layout against a layout where tuples are randomly assigned to MAU stages (called worst case).
We additionally report the average latency of transactions under the two data layouts.
The results are shown in \Cref{fig:exp_partitioning}.

We first discuss the effect of the data layout on the throughput.
For SmallBank, we see that our data layout (called optimal case) has a much more significant effect on the throughput than for the YCSB-A benchmark.
This effect is caused by a higher number of multi-pass transactions for a random data layout in SmallBank that cause an inferior performance.
An optimal data layout, for TPC-C, has almost no effect compared to the random data layout because this workload uses warm transactions where the major limiting factor is the execution of cold sub-transactions rather than the execution of multi-pass transactions on the switch.

If we now look at the latency, we see similar effects.
First, for YCSB and SmallBank, the optimal data layout enables that hot transactions can be executed in a single pass. Hence, the latency remains (almost) stable with increasing load (\# of threads per server). This is in contrast to the random data layout, which causes multi-pass transactions (that require a pipeline-lock) which leads to increased latency under increased load.
For TPC-C, latency increases for both data layouts instead since TPC-C uses warm transactions and here the latency is dominated by the cold sub-transactions.
Overall, the average latencies are low (i.e., in the order of $\mu$s) for all benchmarks.

\vspace{-1.5ex}\subsubsection{Hot-set exceeds Switch's Capacity}
\label{sec:bighotset}
An interesting question is how \system{} performs if the hot-set size exceeds the switch's capacity. 
Hot-sets that exceed the switch capacity can be supported in \system{} by storing a part of the hot-set on normal nodes.
For showing the effect of large hot-sets, we use the YCSB-A workload with increasing hot-set sizes.
Furthermore, we use different tuple-width (i.e., values of different sizes) leading to a lower switch capacity in total number of rows the switch can store.

\begin{figure}
    \captionsetup[subfigure]{aboveskip=0.0ex,belowskip=0.0ex}
    \captionsetup{aboveskip=0.0ex,belowskip=0.0ex}
    \centering
    \includegraphics[width=.99\columnwidth, trim=0mm 0mm 0mm 0mm, clip]{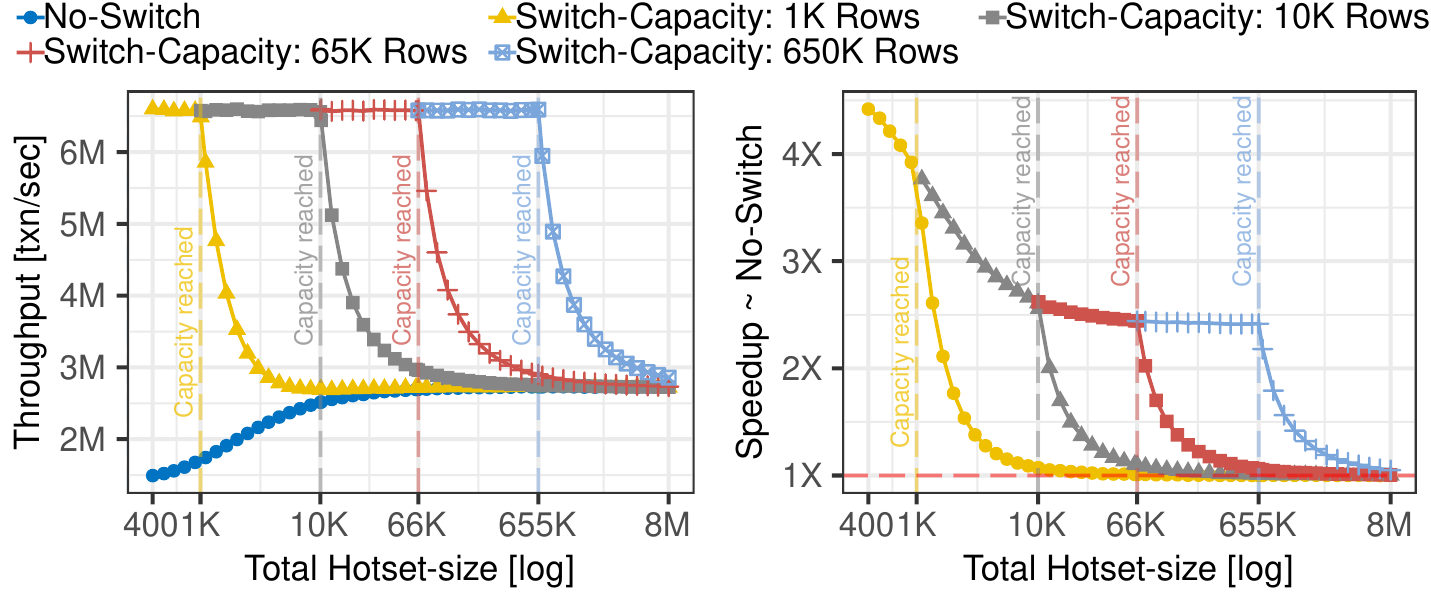}
    \caption{Throughput with growing hot-sets for YCSB-A. Different switch capacities result from different tuple widths. If the hot-set is beyond the switch capacity, the throughput in \system{} degrades gracefully and reaches the throughput of a DBMS without switch support.}
    \label{fig:bighotset}
\end{figure}

\Cref{fig:bighotset} shows how the throughput behaves while growing the hot-set above 4 different fixed switch capacities respectively, the largest being $650$K rows. %
An important fact is that the throughput gracefully degrades when the hot-set outgrows the switch capacity in all cases (Note the log-scale of the x-axis). For example, for the largest switch capacity, \system{} can still achieve approximately $3$ million txn/s for a hot-set of $2$ million tuples which exceeds the switch capacity by more than $2\times$.

\vspace{-1.5ex}\subsubsection{Latency-Breakdown for TPC-C}
\label{sec:tpcc_cycles} 
\begin{figure}
    \captionsetup[subfigure]{aboveskip=0.0ex,belowskip=0.0ex}
    \captionsetup{aboveskip=0.0ex,belowskip=0.0ex}
    \begin{subfigure}{.55\columnwidth}
        \centering
        \includegraphics[width=0.95\linewidth, trim=0mm 0mm 0mm 0mm, clip]{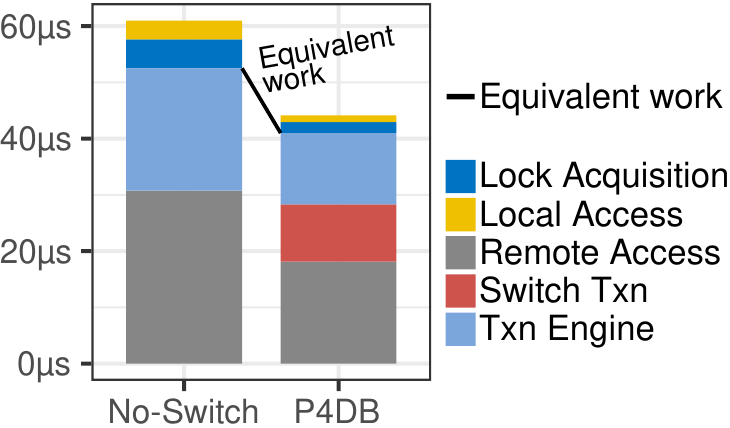}
        \caption{Latency-Breakdown}
        \label{fig:tpcc_cycles}
    \end{subfigure}
    \hfill
    \begin{subfigure}{.39\columnwidth}
        \centering
        \includegraphics[width=0.95\linewidth, trim=0mm 0mm 0mm 0mm, clip]{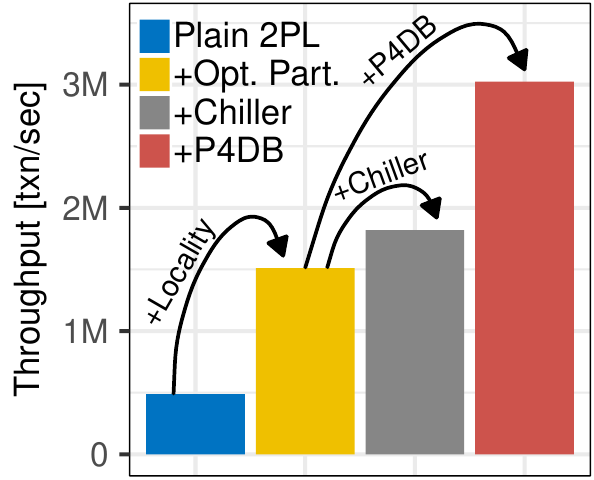}
        \caption{Existing Optimizations}
        \label{fig:tpcc_chiller}
    \end{subfigure}
    \caption{Further Microbenchmarks for TPC-C: (a) Shows the latency-breakdown to understand where the benefits of using \system{} come from. (b) shows effects of existing optimizations for processing distributed and hot transactions.}
    \label{fig:microbench_tpcc}
\end{figure}

In \Cref{sec:eval_tpcc}, we showed that \system{} can achieve a performance gain of more than $2\times$ speed-up for TPC-C under high contention (8 Warehouses).
To understand where the benefits of using \system{} come from, we now show a break-down of the latency for committed transactions for a setup with 8 nodes each with 20 workers. 
The results in \Cref{fig:tpcc_cycles} show the time spent in each component for our baseline without a switch and \system{}. 
Please note, that the total latency shown in this experiment is minimally increased due to profiling overhead compared to the other experiments before.

A first effect visible in \Cref{fig:tpcc_cycles} is that 
\system{} can significantly reduce the time it takes transactions to synchronize on latches to check lock requests compared to the baseline (called \emph{Lock Acquisition}). 
This is due to the execution scheme of \system{} which offloads contended items to the switch (i.e., warehouse, district or hot-stock items for TPC-C).
Since \system{} can execute hot transactions without any locks, the average time spent for \emph{Lock Acquisition} is reduced (i.e., locking is only needed for cold items).
Furthermore, a second effect that leads to the benefits of \system{} is that \emph{remote access} latency of transactions is reduced significantly since remote transactions on hot items only need half a round-trip to the switch instead of a full round-trip to other nodes.
Overall, this results in less time needed to perform the actual work (i.e., remote reads/writes) for remote transactions, as shown in \Cref{fig:tpcc_cycles}.

\vspace{-1.5ex}\subsubsection{Existing Optimizations for Distributed Transactions and Contention.}
\label{sec:tpcc_chiller}

In this experiment, we analyze how existing optimizations for distributed transactions processing and contention compare to \system{}. 
To show the effects of existing optimizations, in this experiment we start with a \emph{Plain 2PL} protocol using two-phase commit (2PC) for distributed transactions without any additional optimizations that clearly results in the lowest throughput.
Afterwards, we add typical optimizations for distributed transactions incrementally to the baseline including \system{}.
We use the TPC-C workload which is the most complex workload from the previous experiments.
Due to space restrictions, we show the results for 8 warehouses only (which results in the highest level of contention). For a higher number of warehouses we see similar results.
The results can be seen in \Cref{fig:tpcc_chiller} compared to \system{}.

As a first optimization, we apply optimal data partitioning to improve data locality of distributed transactions (called \emph{+Opt. Part.} in \Cref{fig:tpcc_chiller}).
In order improve locality in OLTP various partitioning schemes such as \cite{curino2010schism} have been proposed.
For showing the general effect of locality in this experiment, we use two different TPC-C configurations --- one with $80\%$ and another one which results in $20\%$ remote transactions respectively. 
As a second optimization, we additionally apply a contention-centric execution scheme on top of the optimal data partitioning. For the experiment, we added a recent scheme called Chiller \cite{zamanian2020chiller} to our baseline which builds on 2PL but it aims to increase throughput on hot and contented items by deploying a two-region execution scheme with early lock release for contended items instead of using plain 2PL. 
Alternatively to Chiller, we offload the hot-set of hot tuples to the switch and use \system{} for executing transactions on the contended items.
The results show that the optimizations significantly improve the 2PL/2PC protocol but \system{} still outperforms these optimizations clearly.

\section{Related Work}
\label{sec:related_work}

In the following, we discuss related work using INP for OLTP (which is the main target in this paper) in detail. In addition, other related work of using INP for OLAP exists such as \cite{sapio2017daiet,sapio2017network,lerner2019case,blocher2018boosting}.

\vspace*{-1.5ex}\paragraph{INP Support for OLTP}
While INP has been used to support OLTP, to  the best of our knowledge \system{} is the first approach to offload OLTP processing to programmable switches.
\textit{NetCache} \cite{jin2017netcache} proposed a key-value store on a programmable switch to cache hot-items.
Different from this approach, \system{} supports the execution of multiple operations as a transaction directly on the switch, while \textit{NetCache} only supports simple (read/write) operations.
In addition, \textit{NetCache} uses a different approach for reads and writes, which leads to inferior performance of update-heavy workloads.
Another direction is to support replication schemes by INP.
For example, \textit{Harmonia} \cite{zhu2019harmonia} implements a replication protocol for OLTP DBMSs in a switch data-plane to overcome scalability issues.
Moreover, \textit{NetChain} \cite{jin2018netchain} proposes a replication protocol across multiple programmable switches.
All these approaches are clearly orthogonal to \system{}.
Furthermore, another orthogonal work is \cite{jepsennetwork} which proposes an INP approach called Transaction Triaging that manipulates streams of transactions by batching, re-ordering, steering, and protocol conversion of transactions on a programmable switch to improve efficiency.
Finally, \textit{NetLock} \cite{yu2020netlock} proposes a central lock-manager which directly processes lock-requests of hot items in the switch's data-plane.
However, as we have shown in \Cref{sec:eval_switch_lm}, NetLock provides inferior performance for OLTP workloads under high contention when compared to \system{}.

\vspace*{-1.5ex}\paragraph{OLTP under High Contention.}
Another line of work aims to reduce contention with specialized protocols in the host DBMS.
For example, one direction is to reduce contention through enforcing determinism to part of, or all of the concurrency control (CC) unit~\cite{cowling2012granola, kallman2008h, thomson2012calvin}.
Another direction is Quro~\cite{yan2016leveraging}, which re-orders operations inside transactions in a centralized DBMS with 2PL to reduce lock duration of contended data.
While this is relevant in the context of this paper, almost all these works deal with single-node DBMSs and do not have the notion of distributed transactions.
In the context of distributed DBMS, there has recently been some work to leverage high-speed networks and data partitioning to reduce contention in OLTP \cite{zamanian2020chiller}. While this work clearly targets a similar problem, it is orthogonal and thus an interesting direction would be to combine such an approach with \system{} in the future.

\vspace*{-1.5ex}\paragraph{SmartNICs for OLTP}
\label{sec:smartnic}
Previous work leveraged SmartNICs as co-processors to offload distributed transaction processing \cite{schuh2021xenic,liu2019offloading}. %
However, processing hot data on SmartNICs has several drawbacks compared to programmable switches. First, programmable switches are connected to all network nodes centrally and are not bandwidth-limited like SmartNICs with only a few network ports (e.g. 2x50GbE). Processing at full aggregated network-bandwidth is enabled by \system{}'s pipelined and abort-free transaction execution engine implemented on the switch ASIC. 
Second, and more importantly in the OLTP context, programmable switches can be reached in only ½ network latency compared to remote SmartNICs which require a full roundtrip to a node which hosts the SmartNIC. 
While certain SmartNICs that use general-purpose processors for INP allow more complex per-packet processing, they thus also require coordination of data-accesses by multiple threads (e.g. through locking, similar to normal databases nodes) which presents a major bottleneck for hot items as we have shown. Furthermore, upcoming switch models provide increasing resources for computation and memory \cite{inteltofino1,inteltofino2} or even FPGAs as on-chip co-processors \cite{apsnetworks_fpga}.

\section{Conclusion}
\label{sec:conclusion}

In this paper we presented \system{}, a first-of-its-kind approach that leverages a programmable switch to accelerate highly contented OLTP workloads by offloading transaction processing to the switch.
For this, \system{} exposes the switch as an additional database node that not only stores hot tuples, but also implements pipelined and abort-free transaction processing directly in the network's data plane.
The two major benefits of our in-switch design are (1) access in only ½ latency and (2) aggregated processing bandwidth of all connected nodes.
In our evaluation, we have shown that \system{} can thus effectively mitigate contention, i.e., skew on hot tuples, yielding a substantial increase in throughput due to its pipelined transaction execution on a switch. %

\begin{acks}
This work was partially funded by the German Research Foundation (DFG) under the grants BI2011/1 \& BI2011/2 (DFG priority program 2037) and the DFG Collaborative Research Center 1053 (MAKI).
We thank Intel for their valuable technical support.
\end{acks}
\newpage{}

\onecolumn
\begin{multicols}{2}
\bibliographystyle{abbrv}
\bibliography{bibliography.bib}
\end{multicols}
\twocolumn

\clearpage{}
\appendix
\section{Extended Technical Report}

\subsection{Raw Throughput Numbers for Experiments}

\begin{figure*}%
    \begin{subfigure}{.99\columnwidth}
        \centering
        \includegraphics[width=0.99\linewidth, trim=0mm 0mm 0mm 0mm, clip]{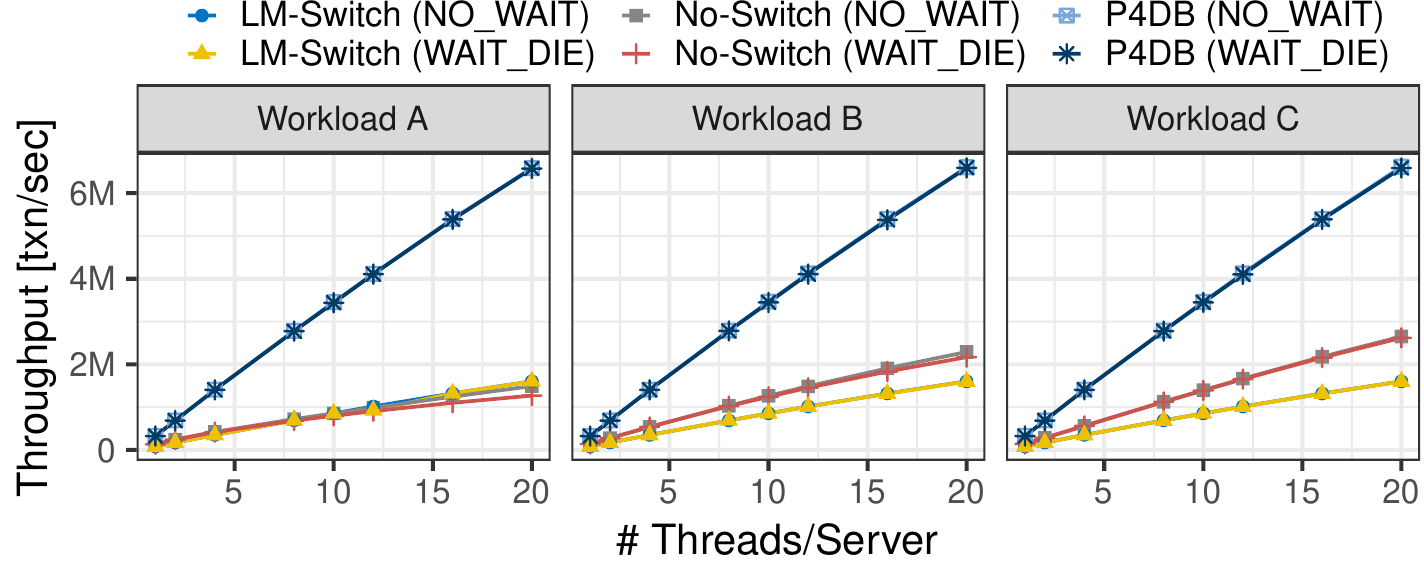}
    \end{subfigure}
    \begin{subfigure}{.99\columnwidth}
        \centering
        \includegraphics[width=0.99\linewidth, trim=0mm 0mm 0mm 0mm, clip]{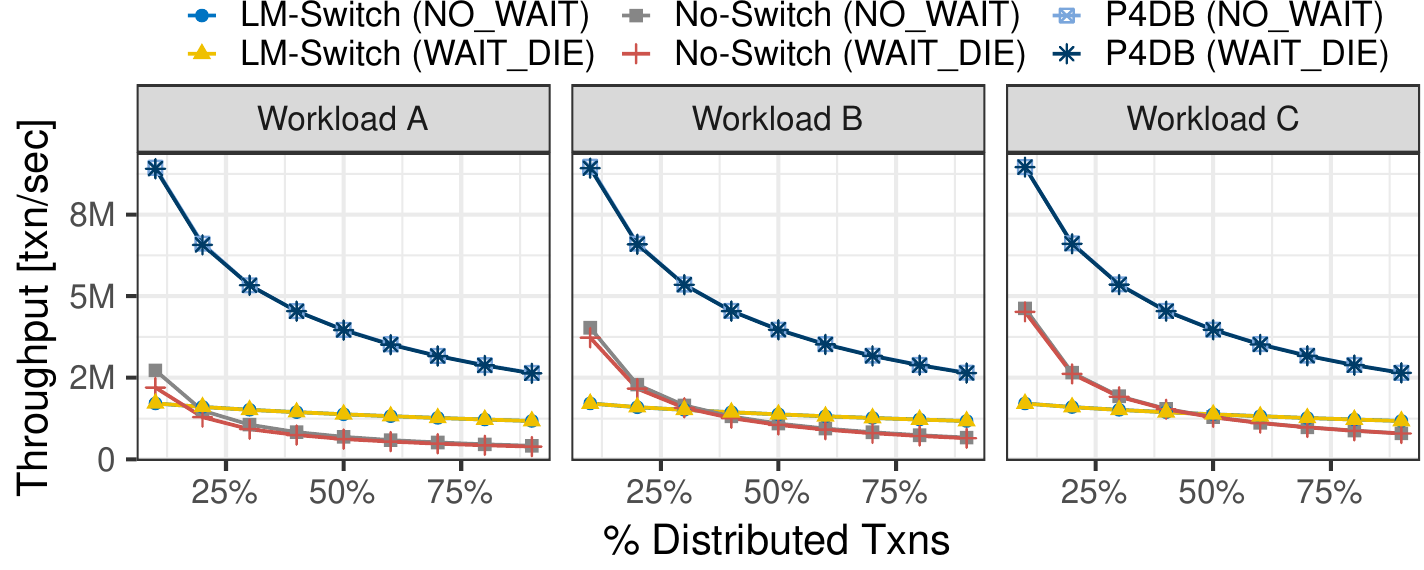} %
    \end{subfigure}
    \vspace{-2.5ex}
    \caption{YCSB Results --- Raw throughput}
    \label{fig:ycsb_raw}
\end{figure*}

\begin{figure*}%
    \begin{subfigure}{.99\columnwidth}
        \centering
        \includegraphics[width=.99\columnwidth, trim=0mm 0mm 0mm 0mm, clip]{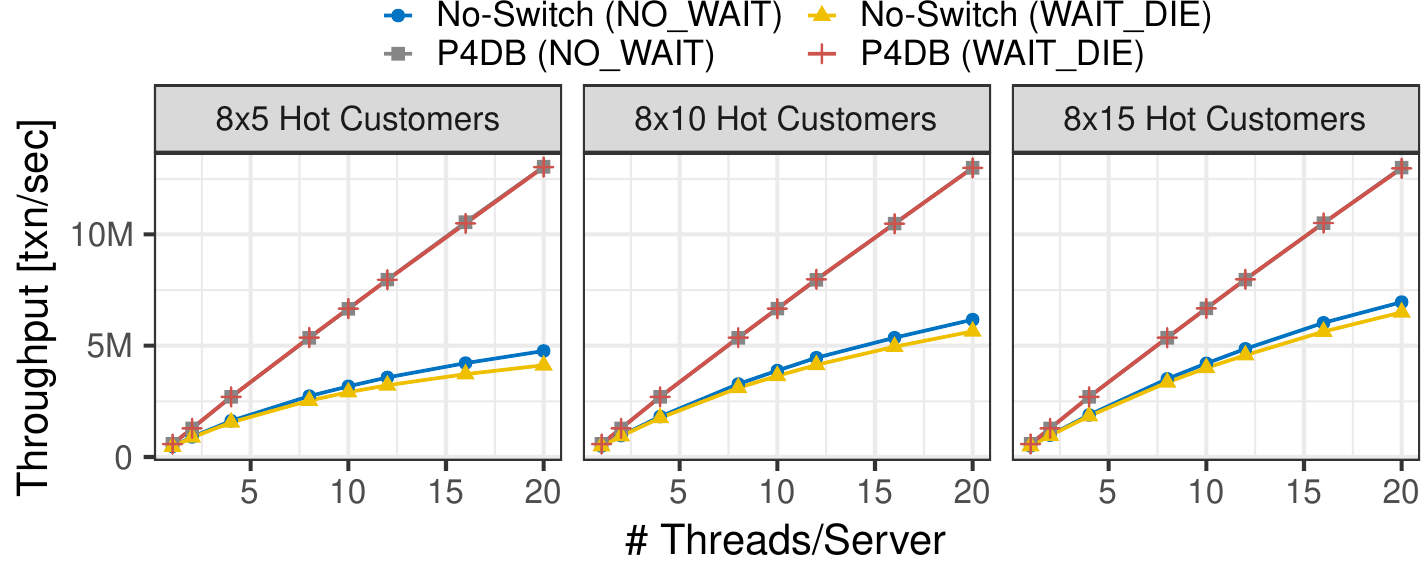}
        \phantomsubcaption
        \label{fig:smallbank_scaleout_raw}
    \end{subfigure}
    \begin{subfigure}{.99\columnwidth}
        \centering
        \includegraphics[width=0.99\linewidth, trim=0mm 0mm 0mm 0mm, clip]{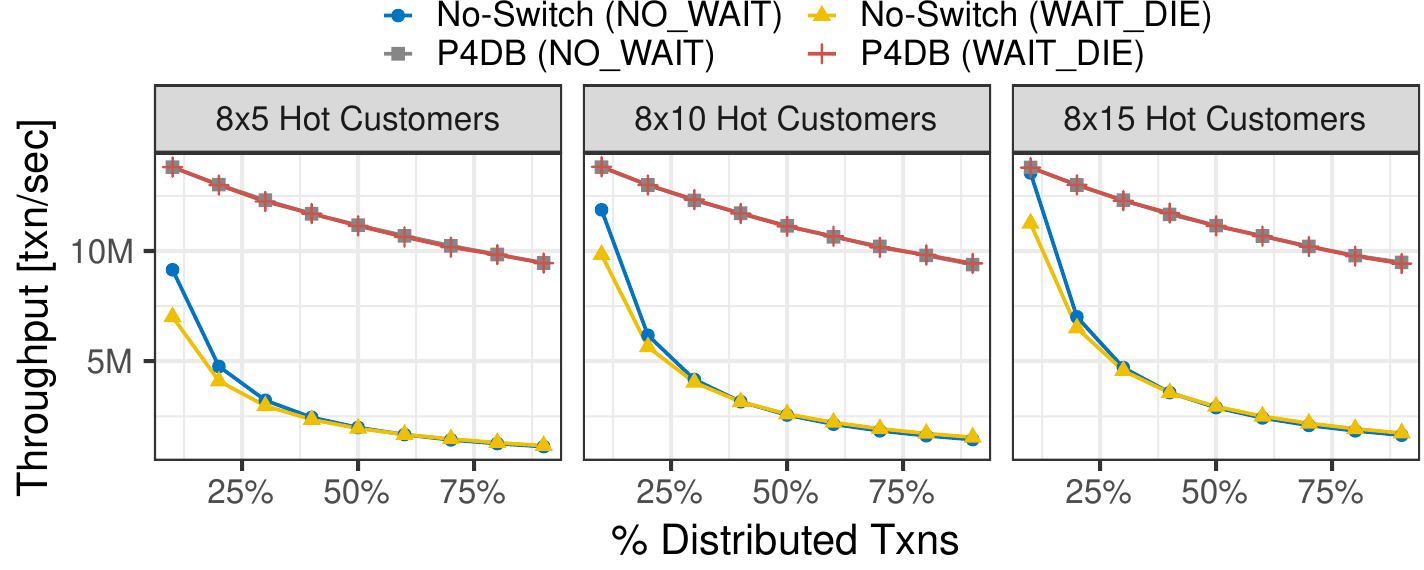} %
        \phantomsubcaption
        \label{fig:smallbank_distributed_raw}
    \end{subfigure}
    \vspace{-2.5ex}
    \caption{SmallBank Results --- Raw throughput}
    \label{fig:smallbank_raw}
\end{figure*}

\begin{figure*}%
    \begin{subfigure}{.99\columnwidth}
        \centering
        \includegraphics[width=0.99\linewidth, trim=0mm 0mm 0mm 0mm, clip]{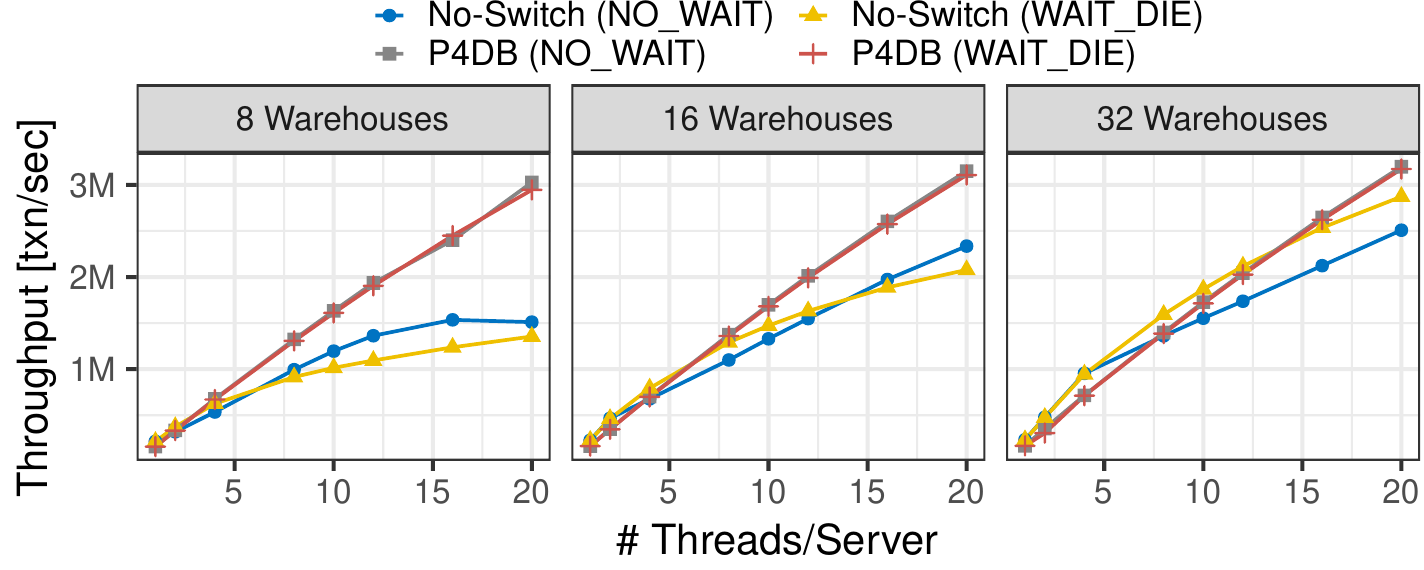}
        \phantomsubcaption
        \label{fig:tpcc_scaleout_raw}
    \end{subfigure}
    \begin{subfigure}{.99\columnwidth}
        \centering
        \includegraphics[width=0.99\linewidth, trim=0mm 0mm 0mm 0mm, clip]{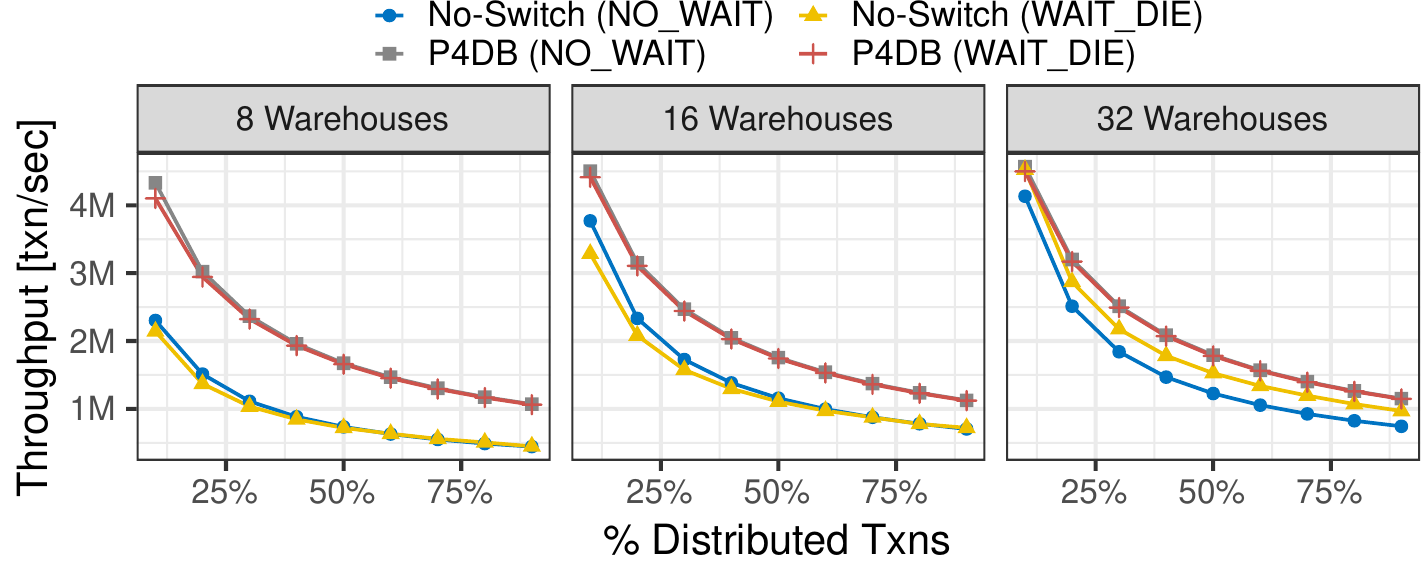} %
        \phantomsubcaption
        \label{fig:tpcc_distributed_raw}
    \end{subfigure}
    \vspace{-2.5ex}
    \caption{TPC-C Results --- Raw throughput}
    \label{fig:tpcc_raw}
\end{figure*}

For the main part of the paper, we opted to only show the speedup numbers between the No-Switch baseline and \system{} for easier comparison and due to space constraints.
In this section we additionally present  plots showing raw throughput numbers (\Cref{fig:ycsb_raw}, \Cref{fig:smallbank_raw} and \Cref{fig:tpcc_raw}) for our experiments. We used these results as a base to calculate the speedup for the plots shown in \Cref{sec:evaluation}.

\subsection{P4 Programming Constraints}

\begin{table*}[]
\small
\begin{tabular}{p{0.25\linewidth}|p{0.65\linewidth}}
\hline
\textbf{Constraints for \pfour{} Programs} & \textbf{Implementation in \system{}} \\ \hline
Control-flow & Different paths in the control-flow can be implement using if-statements or different match-actions entries in tables. \\
Loops & The loop body is unrolled $n$-times at compile-time. If-statements are used if $n$ is a runtime-variable. \\
Floating-point Operations & Fixed point arithmetic, use external FPU if possible. \\
Multiplication/Division & Left/right shifting for values power of two, otherwise decompose or use external ALU if possible. \\
Strings & Use dictionary encoding to process strings. \\
Pointer-based array access & Static allocation and integer-based indexing. Register arrays can be accessed using a calculated index. \\
Dynamic Memory & Redesign algorithm or emulate dynamic allocations using registers and counters. \\ \hline

Limited switch memory & Partition hot data across switch and servers according to \cref{sec:data_layout_problem}. \\
Accesses different than MAU order & Apply Data Layout Algorithm from \cref{sec:data_layout_problem}, fallback to recirculation for exceptions. \\
Read-dependent-writes & Implement as single access atomically using RegisterActions or by conditional statements in the control-flow.\\
Hashing & Use hash-functions provided by Tofino or pass data through checksum-engine. \\
Lookup-Tables & Implement in SRAM or TCAM (Ternary Content-addressable memory) memory. \\ \hline
\end{tabular}
  \caption{Constraints imposed by the \pfour{} programming language on switch programs and how they are handled in \system{}'s implementation. The upper part of the table groups general programming constraints together, whereas the lower part focuses on aspects required for transaction processing.}
  \label{tab:p4_constraints}
  \vspace*{-4.5ex}
\end{table*}

To give an overview on the constraints imposed by \pfour{}, we list the most common ones in \Cref{tab:p4_constraints} and describe how \system{} implements them.
Clearly, the exact capabilities depend on the switch-model, because vendors are free to add accelerated functions through \pfour{}-externs. Thus, we aimed to design \system{} as flexible as possible such that it can be mapped to other switch models easily.

\subsection{Durability - Additional failure cases}

In this section, we discuss how durability and recovery are handled for warm transactions in \system{} in more detail.
Hot transactions (i.e., switch-only transactions) can be handled by the very same procedure.
In general, we need to think about three failure cases: (1) switch fails, (2) nodes involved in a warm transaction fail, (3) both nodes and the switch fail.
In the following, we discuss the cases individually:

\begin{enumerate}
\item The first question that needs to be handled for this case is how a switch-failure can be detected from nodes in \system{}. In general, a switch-failure (e.g., a switch crash) in \system{} can be detected through port-down events on a node. If nodes detect such an event, the recovery process is initiated (once the switch is back).
To enable recovery of the switch state, all committed and pre-committed warm transactions must be considered and the switch state is restored based on the successful switch transactions. 
For enabling recovery of the switch state, database nodes use their local write-ahead log and append the operations for switch transactions they trigger to this log.
In addition, to enable a correct recovery of the switch state from the different local logs of database nodes (in case of a switch failure), the switch adds a globally-unique transaction ID (GID) to each switch transaction that it executes, which represents the (serial) execution order of the transactions on the switch (see: \Cref{sec:durability}).
This ID is sent back together with the result of the read- and write-operations of a switch transaction to the database node in the response packet. The information is then appended to the write-ahead-log of the database node.
In case the switch fails and needs to be recovered, the information in the local logs, which includes the globally ordered transaction IDs of the switch transactions, can then be used to recover a consistent state of the switch. 
A special case to consider for recovery are switch transactions that are in-flight (i.e., they are sent out but the result was not received by nodes). 
In \system{}, it is important to note that switch transactions count as committed before they are sent out since they cannot abort anymore. Therefore, a switch transaction and its intended read-/write-operations are appended to the log before the switch transaction is sent. The only info missing on database nodes at this point in the log is the globally-unique transaction ID (GID) of switch transactions as well as the results of the read/write-operations.
To handle this case, \system{} aims to restore the order from dependencies in the read/write-set (as we discussed in more detail for warm transactions in \Cref{sec:integration_warmextensions} and in the examples below).
If no such dependency is detected, any order of switch transaction can be used during recovery for log replay.

\item The second case is if one or several nodes fail that are involved in a warm transaction (but the switch does not fail). In this case, the DBMS nodes are halted and the crashed nodes need to be recovered. First, the switch state does not need to be recovered since all switch transactions that are sent out are part of warm transactions that are in pre-commit state (and count as committed). For recovering the failed nodes we can use a normal recovery procedure to recover warm transactions (i.e., the writes on cold tuples of all committed and pre-committed warm transactions are recovered while operations on cold tuples of other warm transactions are reverted).
Since the switch-state does not need to be recovered, possible missing transaction IDs (GIDs) of switch transactions in the logs of failed nodes can be ignored. (However, these will be then recovered in case the switch fails.)
A special case is if all nodes that are involved in a warm transaction fail. In this case the nodes can not be sure if the switch also went down, therefore in addition to recovery of the nodes, the switch state is additionally recovered as in case (1) using the log of the nodes.

\item Finally, the last case is the one where both nodes involved in warm transactions and the switch fail. In this case, we first recover the switch from the local logs of the nodes as discussed in case (1). Afterwards, the failed nodes are recovered as discussed in case (2). We discuss also an example as shown in \Cref{fig:durability_recovery_1} for this case below. Further details of this case are explained in \Cref{sec:integration_warmextensions}.
\end{enumerate} %

\vspace{2.0ex}

\noindent In the remaining part of this section, we want to focus on two specific failure scenarios, while applying the procedure from before (see also \Cref{sec:integration}):
\vspace{1.0ex}

\noindent\textbf{Scenario 1} \textit{(Case 3 above). N1 executes T1 on the switch, N2 executes T2 on the switch, N1 and the switch crash. T2 has a data dependency on T1}:\newline
This scenario is visualized in \Cref{fig:durability_recovery_1}. Here, the switch starts with $x{=}1$, while T1 executes $x{+}{=}2$ and T2 executes $x{+}{=}3$. N2 commits T2 successfully and receives $x{=}6$ from the switch as result, even though the switch-results of T1 got lost (i.e., the result of T1 and transaction ID in the log of Node 1 is missing).
In general, this does not present a problem since we can use the read/write-set to restore the order of switch transactions. For the example, we see that for T2 must follow T1 since the local log of T2 indicates that the switch transaction of T1 must be executed before T2; i.e., the initial state is $x{=}1$, afterwards T1 executes $x{+}{=}2$ on the switch, and finally T2 executes executes $x{+}{=}3$.
In case the ordering of two switch transactions would be not deterministic (e.g. both switch-txns executed $x{=}x{+}1$) or different orders are possible, an arbitrary order can be chosen.

\noindent\textbf{Scenario 2} \textit{(Case 2 above). N0 executes T1 on the switch, but crashes before receiving the results. Other transactions depend on T1 and could generate cascading aborts, like in traditional CPU-based execution.}\newline
First, switch transactions are never rolled back and thus cascading aborts caused by switch transactions cannot happen. A switch transaction always commits and the cold part of each transaction on the node needs to be considered as already ``committed''. Therefore, the log entries of  switch transactions need to be written into the log before the switch transaction is sent out. In case one switch transaction is not received by a crashed node, subsequent switch transaction (from other nodes) with dependencies on T1 that committed already are still valid. 
Moreover, for recovery, Node N0 must be recovered as discussed before; i.e., the local log is used to recover all committed transactions on that node and potential gaps of switch transactions are filled in.
However, the switch does not need to be recovered unless all nodes fail (as discussed before).

\noindent
\textbf{Additional remarks:}
On first sight, logging both read- and write-set on the nodes might induce additional overhead. However, this represents no bottleneck in \system{} since each worker can log its own switch transactions independently. This overhead is also negligible in the context of network latency and more importantly when considering the benefits that arise from reduced contention, the overhead clearly amortizes.

\subsection{Other Concurrency Schemes}
\label{res:other_cc_schemes}
In \Cref{sec:integration_warmextensions} of the paper, we discussed how the execution of warm transactions in \system{} is integrated into the 2PC protocol of the host DBMS. This section explains more in-depth how this integration can be done for major classes of concurrency control schemes.

\paragraph{Timestamp-based CC}
This family of CC schemes relies on assigning unique timestamps to transactions and using them for ordering and deadlock prevention. In \textit{TIMESTAMP} \cite{bernstein1981concurrency} a transaction's timestamp dictates whether access to a record owned by another transaction is allowed. If the owning transaction's timestamp is younger than the timestamp from the requesting transaction it is aborted. \textit{MVCC} \cite{bernstein1981concurrency} additionally stores multiple versions of records for different timestamps instead of only one. %
Both variants typically contain a validation phase before commit. Here, warm transactions can be integrated similarly to \system{} by issuing the switch sub-transaction on the hot items between the validation and commit phase, as this ensures that the warm transaction can not abort anymore.

\paragraph{Optimistic CC}
In optimistic concurrency control (\textit{OCC}) transactions are executed concurrently and before commit it is verified whether the execution was in fact serializable or not. The DBMS validates this against all other transactions that committed or are currently in the validation phase, since the transaction started. Here, similar as before, a coordinator collects commit or abort decisions from all participating servers and commits the transaction if every server voted to commit. Thus, to integrate warm transactions, the coordinator sends and receives the switch sub-transaction on the hot items before broadcasting the commit-decision.

\paragraph{Deterministic CC}
In deterministic CC, e.g. \textit{CALVIN} \cite{thomson2012calvin,harding2017evaluation}, centralized coordinators decide on a deterministic order of transactions that is abort-free and thus no coordination between servers is required like in other protocols.
For this protocol, while hot transactions can be ignored in the ordering (since the switch guarantees abort-free execution), warm transactions are included in the ordering and then can be executed without further coordination in the given order including the switch sub-transaction on the hot items.

\subsection{Other Deployments}
\label{sec:res_deployments}

While \system{} focuses only on a single-tenant DBMS deployment in a rack with a top-of-the rack switch, other deployments are also conceivable, which we briefly sketch in the following.

\paragraph{Switch Hierarchies.}
Data center networks that provide connectivity between nodes often build on hierarchies with multiple switches: The lowest layer consisting of top-of-the-rack (ToR) switches bundles a subset of nodes together to ensure low-latency communication and high throughput, while multiple ToR switches are then cross-connected in subsequent layers to provide connectivity to all nodes.
Using such switch hierarchies, \system{} can make use of the information about the topology to improve performance. For example, the nearest programmable switch running \system{} could be used for offloading to avoid sending packets through a long path consisting of multiple hops.
Another use-case could be to utilize multiple switches for replication, to ensure that the system is robust against failures by leveraging protocols such as \cite{jin2018netchain}.

\paragraph{Multi-Tenancy}
Multi-tenancy means in our context that a single instance of \system{}  serves multiple customers. Such a multi-tenancy setup for different customers is actually covered already by benchmarks such as SmallBank or TPC-C that we include in our evaluation in Section \ref{sec:evaluation}.
While tenants typically share the same database schema in such a setup, in addition, data of each tenant needs to be isolated to remain invisible to other tenants. This is generally possible in \system{} by partitioning the switch resources, thus making it impossible to access or modify data from other tenants.
However, the limited switch resources then still need to be kept in mind when deciding which hot tuples are offloaded to the switch.

Finally, we actually see also some potentials by including multi-tenancy in \system{}. Instead of using separate register arrays for each tenant, a data layout on the switch which spreads the data of each tenant across as many register arrays as possible is beneficial, because the amount of access conflicts is reduced. Flexibility is also given for the use of hot data across multiple tenants, e.g. one tenant generates 90\% of the load, whereas other tenants are barely active.

\end{document}